\documentclass[aps,prc,twocolumn,reprint]{revtex4}

\usepackage{graphicx} \usepackage{amsmath,amssymb,amsopn}
\usepackage{latexsym} \usepackage{bm} \usepackage{mathrsfs}
\usepackage{mathbbol} \usepackage[dvips]{color}

\begin{document}
\title{Bottomonia in the Quark-Gluon Plasma and their Production at 
RHIC and LHC}

\author{A.~Emerick} 
\affiliation{School of Physics and Astronomy, University of Minnesota, Minneapolis, MN 55414, USA} 
\author{X.~Zhao} 
\affiliation{Department of Physics and Astronomy, Iowa State University, 
Ames, IA 50011, USA}
\author{R.~Rapp}
\affiliation{Cyclotron Institute and Department of
Physics\&Astronomy, Texas  A\&M University, College Station, TX 77843-3366, USA}

\date{\today}

\begin{abstract}
We study the production of bottomonium states in heavy-ion reactions at 
collider energies available at RHIC and LHC. We employ an earlier 
constructed rate equation approach which accounts for both suppression 
and regeneration mechanisms in the quark-gluon plasma (QGP) 
and hadronization phases of the evolving thermal medium. Our previous 
predictions utilizing two limiting cases of strong and weak bottomonium
binding in the QGP are updated by (i) checking the compatibility of the
pertinent spectral functions with lattice-QCD results for euclidean
correlators, (ii) adapting the initial conditions of the rate equation
by updating bottom-related input cross sections and the charged-particle
multiplicity of the fireball, and (iii) converting our calculations 
into observables as recently measured by the STAR and CMS experiments. 
Our main findings are a preference for strong $\Upsilon$ binding as
well as a significant regeneration component at the LHC.
\end{abstract}

\maketitle
%%%%%%%%%%%%%%%%%%%%%%
\section{Introduction}
\label{sec:intro}
%%%%%%%%%%%%%%%%%%%%%%
Quarkonia by now have a 25-year history of being utilized as a probe  
of the Quark-Gluon Plasma (QGP) in ultrarelativistic 
heavy-ion collisions (URHICs), 
cf.~Refs.~\cite{Rapp:2008tf,Kluberg:2009wc,BraunMunzinger:2009ih} for
recent reviews. The majority of the studies to date have been 
devoted to charmonium properties and their observable consequences in 
experiment, with few exceptions for 
bottomonia~\cite{GV97,Pal:2000zm,Goncalves:2001vn,Grandchamp:2005yw,Andronic:2006ky,Liu:2010ej}.   
Recently, experiments at the Large Hadron Collider 
(LHC)~\cite{Chatrchyan:2011pe,cms-11} and the Relativistic Heavy-Ion 
Collider (RHIC)~\cite{star-11} have measured $\Upsilon$ mesons 
in URHICs for the first time, and subsequent phenomenological analyses 
have commenced~\cite{Strickland:2011mw,Brezinski:2011ju}, focusing on 
suppression mechanisms in Pb-Pb($\sqrt{s}$=2.76\,ATeV) in comparison 
to CMS data. For charmonia, regeneration effects due to coalescence of 
$c$ and $\bar c$ quarks have complicated the interpretation of 
observables appreciably. Since the cross section for open bottom 
is much smaller than for open charm, regeneration of bottomonia 
has been argued to be negligible. 
This would render bottomonium production in URHICs a more pristine 
probe of quarkonium dissolution in the QGP. However, quantitative 
estimates~\cite{Grandchamp:2005yw,Andronic:2006ky} indicate that this 
may not be the case. The magnitude of the regeneration contribution
depends on a subtle interplay of the masses of the open- and 
hidden-bottom states in a system with fixed $b\bar b$ content.
In addition, the ratio of hidden- to open-bottom states in elementary
hadronic collisions is typically of order 0.1\% -- and thus much smaller
than in the charm sector where it is around 1\% --, implying that even
small contributions to bottomonium regeneration can be significant
relative to primordial production. Furthermore, a consistent 
interpretation of bottomonium production in URHICs should encompass
both LHC and RHIC data. 

The above aspects were addressed in previous 
work~\cite{Grandchamp:2005yw} where predictions for the nuclear 
modification factor, $R_{AA}$, of inclusive $\Upsilon$ production at 
RHIC and LHC have been presented. Unfortunately, these calculations 
could not be directly mapped to the now available data: at RHIC the 
$R_{AA}$ for the sum of 1S, 2S and 3S states has been measured, and
at LHC the current data are taken at half of the design energy, 
$\sqrt{s}$=2.76\,ATeV. 
Furthermore, the availability of bulk-hadron observables in Pb-Pb 
collisions at the LHC helps to constrain the fireball model in terms 
of its total entropy and transverse expansion which govern the 
temperature evolution, thus affecting bottomonium suppression and 
regeneration. Calculations of the latter also benefit from an updated 
input for the bottom(onium) cross sections (if only suppression is
considered, this input cancels out in the suppression factor). We 
also improve the theoretical input to the earlier calculations by 
applying theoretical constraints on the in-medium bottomonium 
properties (binding energy and dissociation width) in terms of 
euclidean correlators computed in recent years within thermal 
lattice-QCD (lQCD)~\cite{Jakovac:2006sf,Aarts:2010ek,Aarts:2011sm}. In 
Ref.~\cite{Grandchamp:2005yw}, two scenarios for in-medium bottomonia 
were discussed, i.e., one with vacuum and one with in-medium binding
energies estimated from a screened Cornell potential~\cite{Karsch:1987pv}.
With the now available lattice data, we can refine the pertinent spectral 
functions following the approach outlined in Ref.~\cite{Zhao:2010nk}.

Our paper is organized as follows. 
In Sec.~\ref{sec:sf} we revisit the spectral properties of bottomonia
in an equilibrated QGP within two limiting scenarios, ``strong" and 
``weak" binding, and check the resulting euclidean correlators 
against lQCD data. 
In Sec.~\ref{sec:kin} we briefly recall the kinetic rate-equation 
framework in a thermal medium and discuss its (suitably
updated) input quantities.
In Sec.~\ref{sec:exp} we present the updated results on bottomoniun 
production at RHIC and LHC, focusing on its centrality dependence in 
comparison to STAR and CMS data. We evaluate our results and put them 
into context with existing calculations. 
In Sec.~\ref{sec:concl} we summarize and conclude.

%%%%%%%%%%%%%%%%%%%%%%%%%%%%%%%%%%%%%%%%%%%%%%%%%%%%%%%%%%%%%%%%%%%%%
\section{Bottomonium Spectral Functions in the Quark-Gluon Plasma}
\label{sec:sf}
%%%%%%%%%%%%%%%%%%%%%%%%%%%%%%%%%%%%%%%%%%%%%%%%%%%%%%%%%%%%%%%%%%%%%%
Following Ref.~\cite{Grandchamp:2005yw} we attempt to bracket the 
properties of bottomonia in the QGP by defining two scenarios for the 
temperature dependence of their binding energy,
\begin{equation}
\label{E_B}
\epsilon_B(T)=2m_b(T)-m_Y \ 
\end{equation}
($m_b(T)$: in-medium bottom mass; $m_Y$: bottomonium bound-state
mass with $Y=\Upsilon$(1S), $\Upsilon$(2S), $\chi_b$(1P), $\dots$ ).
These are:
\begin{enumerate}
\item Weak Binding Scenario (WBS): bottomonium bound-state energies are 
significantly reduced in the QGP, relative to the vacuum state, adopting 
the screened Cornell-potential results of Ref.~\cite{Karsch:1987pv}.
The bound state masses are kept at their vacuum value which is 
motivated by the approximate constancy of their in-medium correlator
ratios as found in lQCD (further details below). This, in turn,
requires the in-medium bottom quark-threshold to decrease with
increasing temperature (which is not inconsistent with independent
lQCD computations of the heavy-quark free energy). 
\item Strong Binding Scenario (SBS): the temperature dependence is 
removed from Eq.~\eqref{E_B}, implying bound states to remain at their
vacuum mass and binding energy throughout the QGP, with a constant
open-bottom threshold at $2m_b=2m_B = 10.56$\,GeV. 
\end{enumerate}
\begin{figure}[!t]
   \centering
   \includegraphics[width=0.95\linewidth,clip=]{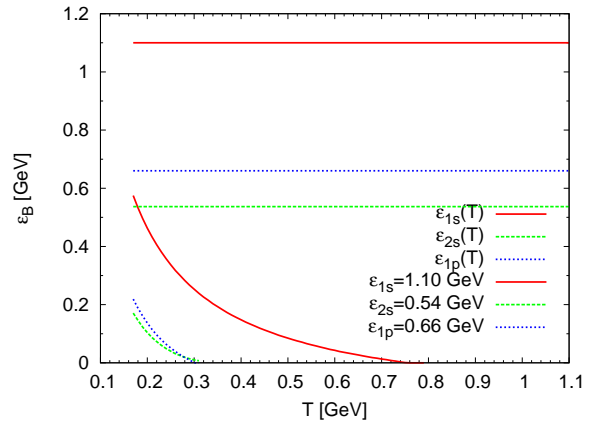}
   \caption{Temperature dependence of the binding energies, $\epsilon_B(T)$,
            for $\Upsilon$ (solid lines), $\Upsilon^{\prime}$ (dotted
            lines) and $\chi_b$ (dashed lines) in the Strong- and
            Weak-Binding scenarios (upper and lower 3
            lines, respectively).}
        \label{fig:E_B}
 \end{figure}
Figure~\ref{fig:E_B} shows the temperature dependences of $\epsilon_B$ 
for the ground state $\Upsilon$, for $\Upsilon^\prime$ and 
$\chi_b$(1P), within these two scenarios. The constant values in the 
SBS are to be contrasted with factor of at least 2 reduction at 
$T_c$=180\,MeV (and rapidly decreasing further with $T$) in the WBS.
Strong- and weak-binding scenarios also emerge in potential-model
calculations, due to current uncertainties in defining the in-medium
interaction kernel, usually bracketed by the internal and free 
energies of the $Q\bar Q$, see, e.g., Ref.~\cite{Riek:2010fk}. 
In the present context we are not (yet) aiming at a quantitative
implementation of potential-model results, but rather at exploring
the sensitivity of heavy-ion data to $Y$ spectral properties within 
the two cases defined above. 

\begin{figure*}[!t]
\includegraphics[width=0.48\linewidth,clip=]{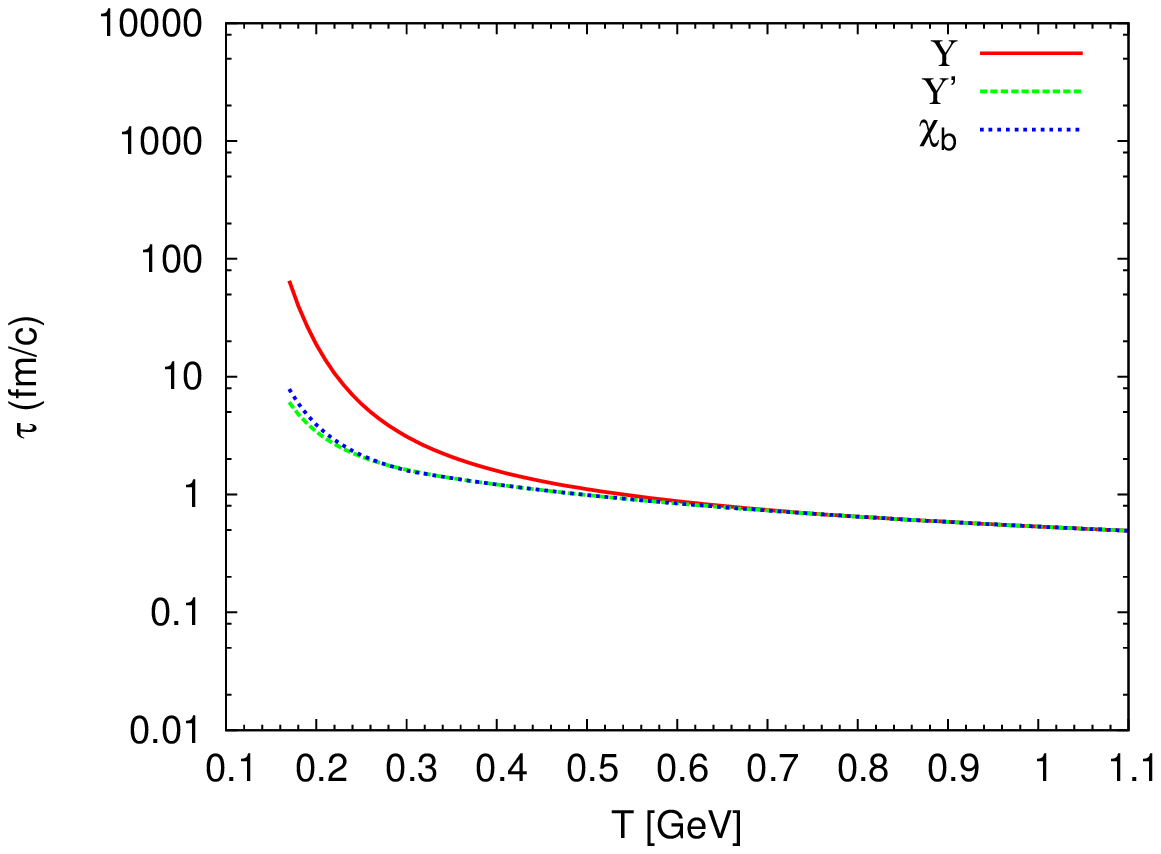}
\includegraphics[width=0.48\linewidth,clip=]{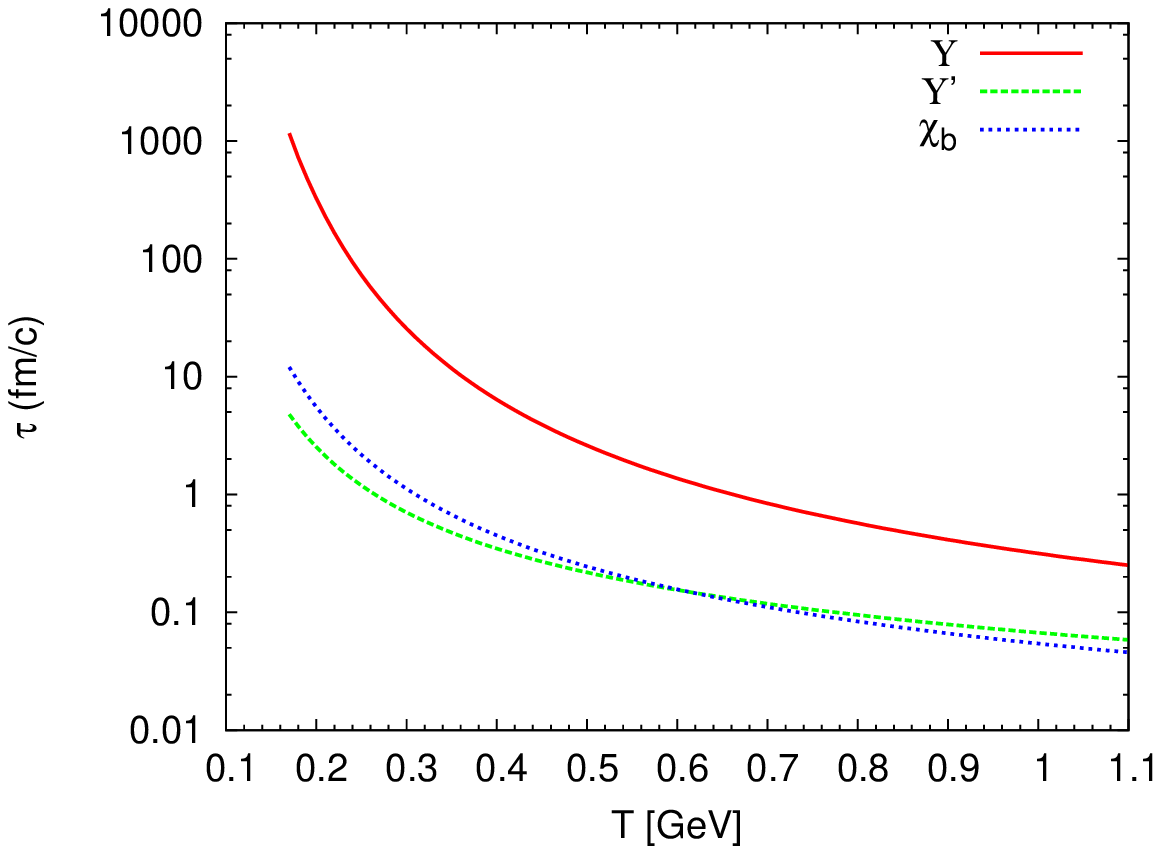}
\caption{Bottomonium lifetimes in the QGP for the two binding scenarios
defined in the text; left panel: WBS with quasifree dissociation; right:
SBS with gluo-dissociation;  solid lines: $\Upsilon$,
dashed lines: $\Upsilon^{\prime}$, dotted lines: $\chi_b$.}
\label{fig:gamma}
\end{figure*}
The next step is the evaluation of the $Y$ dissociation widths. For a 
Coulombic bound state with large binding energy, 
$\epsilon_B\gg \Lambda_{\rm QCD}$, the leading break-up mechanism 
has been identified and calculated more than 30 years ago as the
gluo-dissociation  process,
$g+{\cal Q} \to Q+\bar Q$ (${\cal Q}$: 
quarkonium)~\cite{Peskin79}. The corresponding cross section is peaked 
around incoming gluon energies of $k_0=(10/7) \epsilon_B$. For weakly 
bound quarkonia, $\epsilon_B\le T$, this mechanism becomes inefficient 
and is supplanted by the quasifree break-up, 
$p+{\cal Q} \to p+Q+\bar Q$ ($p=g,q,\bar q$)~\cite{Grandchamp:2001pf}.
This picture has been well established by now. For example, in
effective-field-theory approaches to quarkonia, the gluo-dissociation
corresponds to the so-called ``singlet-to-octet" 
transitions~\cite{Riek:2010py,Brambilla:2011sg} while
the quasifree mechanism corresponds to Landau damping in the 
gluon-exchange potential~\cite{Laine:2006ns,Riek:2010py}. 
Accordingly, we will treat the SBS with the gluo-dissociation and the 
WBS with the quasifree mechanism. The pertinent dissociation rates
(inelastic widths) can be expressed in terms of an underlying cross 
section, $\sigma_Y^{\rm diss}$, convoluted over the thermal parton
distribution functions, $f^p$, as 
\begin{eqnarray}
\label{eq:diss-rate}
\Gamma_Y^{\rm diss}(q;T) &\equiv& \tau_Y^{-1}(q;T) 
\nonumber\\
&=& \sum\limits_{p=q,\bar q,g} \int \frac{\mathrm{d}^3k}{(2\pi)^3} 
f^p(\omega_k;T) \ \sigma_Y^{\rm diss}(s) \ v_{\rm rel} \ , 
\end{eqnarray}
where $\omega_k$ denotes the on-shell energy of a parton with
3-momentum $k$, and $s=(k^{(4)}+q^{(4)})^2$ is the total center-of-mass 
energy squared in the collision with a $Y$ of 3-momentum $q$ 
($v_{\rm rel}$ is their relative velocity). The bottomonium lifetimes 
are plotted in Fig.~\ref{fig:gamma} for $\vec{q}=0$, for $\Upsilon$, 
$\Upsilon^\prime$ and $\chi_b$ in the weak- and strong-binding scenarios
(left and right panel, respectively). As in Ref.~\cite{Grandchamp:2005yw}
quasifree dissociation is calculated with $\alpha_s$$\simeq$0.26, while
for gluo-dissociation a Coulomb-like binding for $\Upsilon$(1S) with
$\epsilon_B$=1.1\,GeV translates into an effective coupling of
$\alpha_s$$\simeq$0.65. Within the WBS, quasifree dissociation induces 
little difference between the $\Upsilon^\prime$ and $\chi_b$ widths, due 
to their almost identical (small) binding energies. The $\Upsilon$(1S) 
lifetime is about an order of magnitude longer than for the excited 
states at $T_c$, but quickly approaches their lifetimes since 
$\epsilon_B$ decreases to small values at higher temperature. In the SBS 
gluo-dissociation produces a larger spread in the lifetimes of ground 
and excited states. At $T_c$, the $\Upsilon$(1S) lifetime is by almost 
two orders of magnitude larger than its WBS counterpart, but it comes 
close to the latter at high temperatures, $T\simeq1$\,GeV. However, it 
stays well above the lifetimes of the excited states even at high $T$. 
Close to $T_c$, the lifetimes of the excited states in the SBS are quite 
comparable to their WBS counterparts, even dropping below the latter for 
$T$$\ge$250\,MeV or so. The reason for that is the larger coupling in 
the SBS together with a favorable overlap of the typical thermal gluon 
energies, $\bar\omega_k\simeq 3T$, with the maximum in the 
gluo-dissociation cross sections around 
$\omega_k$$\sim$$\epsilon_B$$\simeq$600\,MeV. 
Thus, for the phenomenologically relevant temperatures, especially at 
RHIC, one expects the suppression factors of the excited states to be 
similar in the SBS and WBS.

\begin{figure*}[!t]
   \centering
   \includegraphics[width=0.48\linewidth]{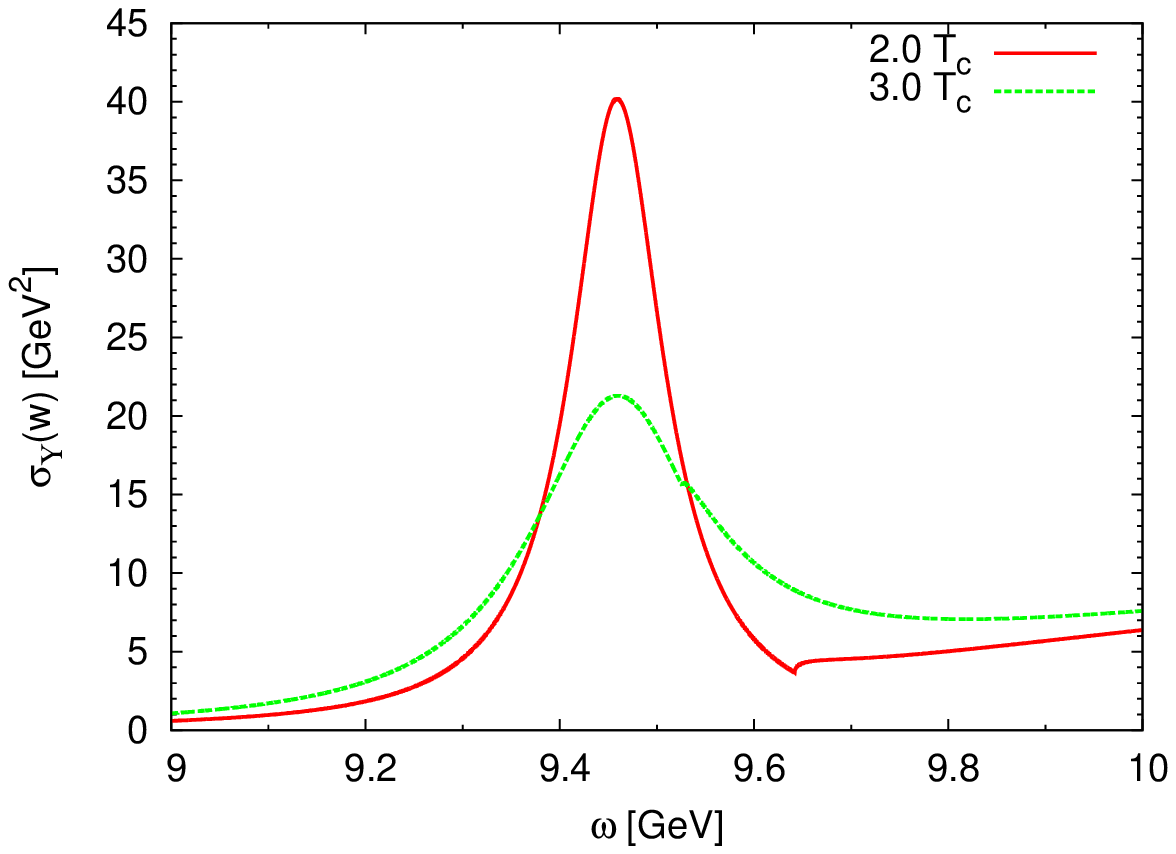}
%   \vspace{0.8cm}
   \includegraphics[width=0.48\linewidth]{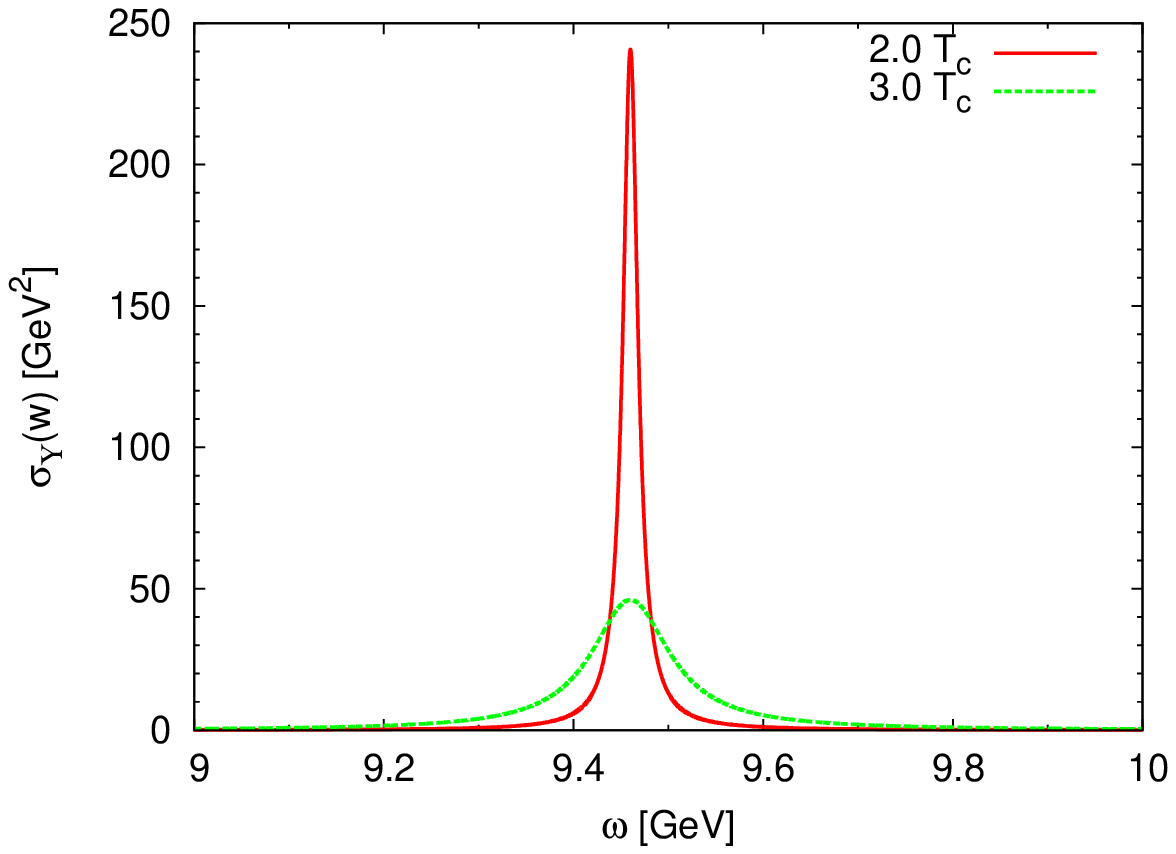}
   \caption{Spectral functions of the $\Upsilon$ calculated from
    Eq.~\eqref{sf} in a QGP of temperatures 2\,$T_c$ (solid lines) 
    and 3\,$T_c$ (dashed lines); left panel: weak-binding scenario,
    right panel: strong binding scenario.}
   \label{fig:sf}
 \end{figure*}

With the masses and widths (real and imaginary parts of the inverse
propagator) of the $Y$ bound states determined we proceed to compute 
the pertinent spectral functions. 
We follow Ref.~\cite{Zhao:2010nk} for the charmonium case by adopting 
an ansatz consisting of a relativistic Breit-Wigner (RBW) for the 
bound-state part and a nonperturbative continuum for energies above 
the $b\bar b$ threshold. We neglect the excited states of $Y$. One has 
\begin{eqnarray}
\label{sf}
\sigma_{Y}(\omega)=A_{Y}\frac{2\omega}{\pi}\frac{Z_Y \omega\Gamma_{Y}(T)}
 {(\omega^{2}-m_{Y}^{2})^{2}+\omega^{2}\Gamma_{Y}(T)^{2}} \qquad \qquad
\nonumber\\
+ \frac{B_{Y}N_{c}}{8\pi^{2}}\Theta(\omega^2-s(T))\omega^{2}
\sqrt{1-\frac{s(T)}{\omega^{2}}}\left(a+b\frac{s(T)}{\omega^{2}}
\right) \ , 
\end{eqnarray}
where $N_c$=3 is the number of colors, $A_{Y}$ is given by the 
wave function overlap at the origin in vacuum, $Z_Y$ characterizes 
its modification in medium (``polestrength"), $s(T)=2m_b(T)$ marks 
the in-medium $b\bar b$ threshold following
Eq.~(\ref{E_B}), and the coefficients $(a,b)$ arise from the Dirac 
structure specific to the hadronic channel under consideration, 
e.g., (2,1) for the vector channel. The prefactor $B_{Y}=2$ augments 
the perturbative expression of the continuum to account for $b\bar b$ 
rescattering effects and is estimated from the $T$-matrix calculations 
of Ref.~\cite{Riek:2010fk}. We approximate the total width for the 
bound state with its dissociation rate (inelastic width). 
Figure~\ref{fig:sf} shows the 
in-medium $\Upsilon$ spectral function for two temperatures in the 
WBS and SBS. One clearly recognizes the much larger width and reduced 
threshold in the WBS compared to the SBS. In the latter the $b\bar b$ 
threshold remains constant at $\sqrt{s}=10.56$\,GeV.

\begin{figure*}[!t]
    \centering
    \includegraphics[width=0.48\linewidth]{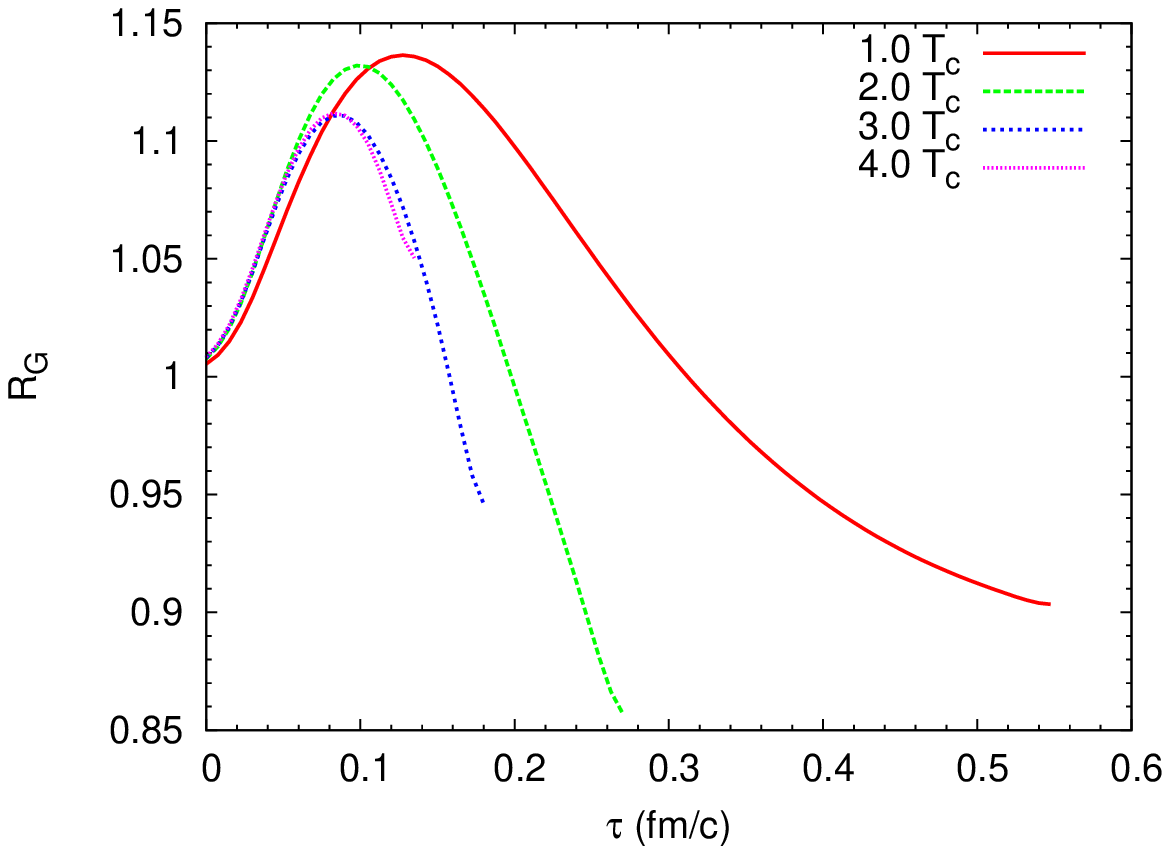}
%    \vspace{0.8cm}
    \includegraphics[width=0.48\linewidth]{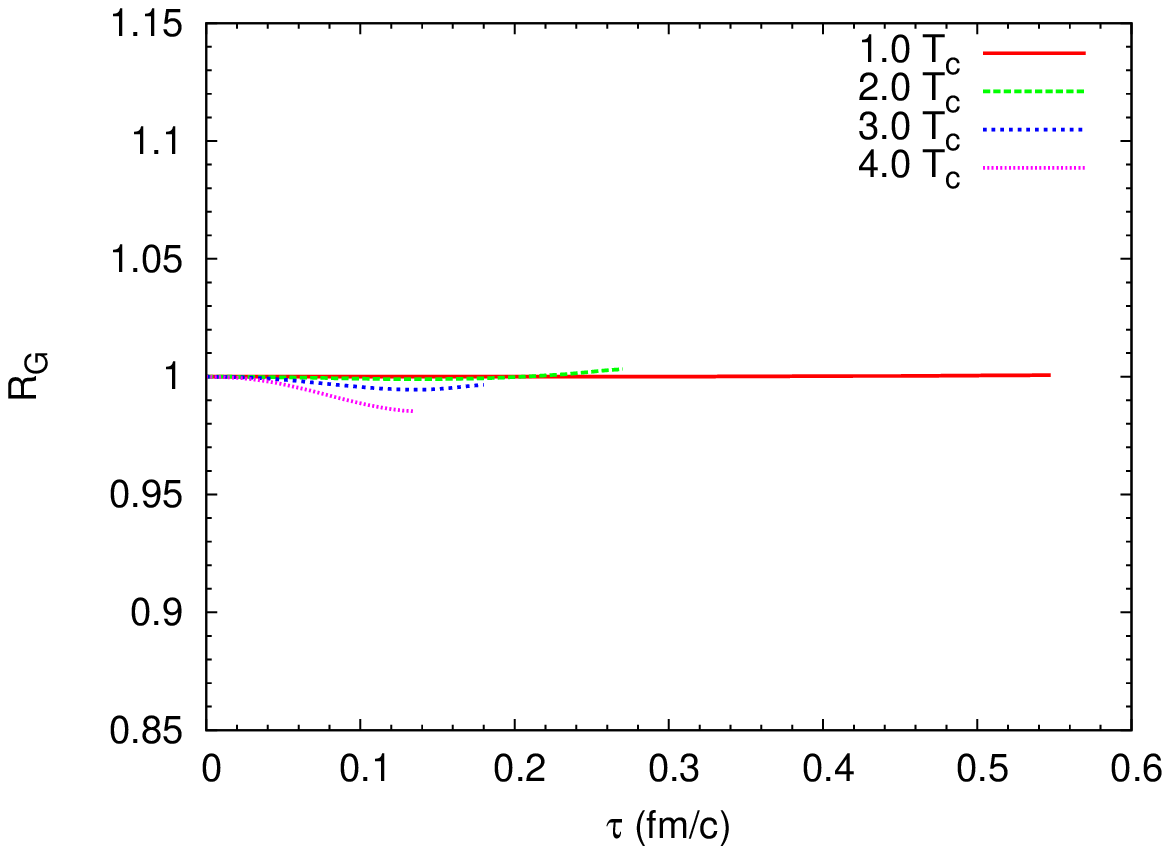}
\caption{Euclidean correlator ratios following from the in-medium
$\Upsilon$ spectral functions shown in Fig.~\ref{fig:sf} at various
temperatures for the weak-binding (left panel) and strong-binding
scenario (right panel).}
\label{fig:ecorr}
\end{figure*}
In a final step we check the above spectral function against results
from thermal lQCD where the bottomonium correlators, $G(\tau)$, are
computed as a function of euclidean time $\tau$. The euclidean
correlator is related to the spectral function as 
\begin{equation}
\label{eq:temp-corr}
G_\alpha(\tau;T)=\int\limits_0^\infty {\mathrm{d}\omega} 
 \ \sigma_\alpha(\omega;T) \ K(\omega,\tau;T) \ , 
\end{equation}
where $\alpha$ specifies the hadronic channel, and the kernel reads
\begin{equation}
\label{eq:kernel}
K(\omega,\tau;T)=\frac{\cosh\left[\omega(\tau-1/2T)\right]}
{\sinh\left[\omega/2T\right]}  \ .
\end{equation}
To better exhibit the medium effects, it is common to consider the
correlator ratio,
\begin{equation}
\label{eq:corr-ratio}
R_G(\tau;T)=\frac{G_\alpha(\tau,T)}{G_\alpha^{rec}(\tau,T)}  \ ,
\end{equation}
where the denominator is evaluated with the same kernel but with
a spectral function evaluated at a low (or zero) temperature, 
$T^*$, where no significant medium effects are expected. 
To compute the euclidean correlator ratios for our in-medium spectral
functions we define a vacuum spectral function using Eq.~\eqref{sf} 
in the narrow-width limit and with $Z_Y$=1 for the RBW part and with 
$\sqrt{s}=2m_B=10.56$\,GeV for the continuum (as for the in-medium
spectral function, we only include the ground state).  Our results for 
$R_G(\tau,T)$ are displayed in Fig.~\ref{fig:ecorr}. In the 
strong-binding scenario (featuring vacuum binding energies and $b\bar b$ 
threshold), the only medium effect is the increasing decay width, which, 
however, has very little impact on the correlator ratio so that the 
latter stays within 2\% of one up to the highest temperature considered 
(4\,$T_c$), cf.~right panel of Fig.~\ref{fig:ecorr}. In the weak-binding 
scenario, the bound-state mass is also constant, but the threshold 
energy drops appreciably, which by itself would imply an increase of 
$R(\tau)$ above one. However, a weakening bound state is not only
characterized by a reduction in $\epsilon_B$ but also by a reduction of
its polestrength (see, e.g., Ref.~\cite{Riek:2010fk}), which is 
represented by the $Z_Y$-factor in the RBW. We have tuned this quantity 
to recover $R(\tau)$ at around one resulting in the curves shown in the 
left panel of Fig.~\ref{fig:ecorr}. The deviations from one are larger 
in the WBS compared to the SBS, comparable to what has been found for 
charmonia with the same ansatz~\cite{Zhao:2010nk}. The SBS, therefore, 
appears to be a more natural realization of the near-constancy 
of $R(\tau)\simeq1$ at the few-percent level as found 
in thermal lQCD~\cite{Jakovac:2006sf,Aarts:2010ek,Aarts:2011sm}.
This insight is one of the main motivations for re-evaluating the 
previous phenomenological analysis of Ref.~\cite{Grandchamp:2005yw} 
where the in-medium binding scenario with quasifree dissociation was 
believed to be more realistic for bottomonia in the QGP. To our 
knowledge, this has not been done in any other phenomenological 
application thus far.

%%%%%%%%%%%%%%%%%%%%%%%%%%%%%%
\section{Kinetic-Theory Model}
\label{sec:kin}
%%%%%%%%%%%%%%%%%%%%%%%%%%%%%%
In this section we first set up the rate-equation framework
for the $Y$ states in the expanding QGP (Sec.~\ref{ssec_rate}),
followed by specifying the inputs and initial conditions for their 
numerical evaluation, specifically
\begin{itemize}
\item[(i)] the temperature and volume evolution of the thermal bulk 
          medium (Sec.~\ref{ssec_fb}), and
\item[(ii)] the initial conditions for the various $Y$ states
            and the number of $b\bar b$ pairs in the system
           (Sec.~\ref{ssec_xsec}).
\end{itemize}

%%%%%%%%%%%%%%%%%%%%%%%%%%%%%%
\subsection{Rate Equation}
\label{ssec_rate}
\begin{figure}[htbp]
\centering
   \includegraphics[width=0.95\linewidth,clip=]{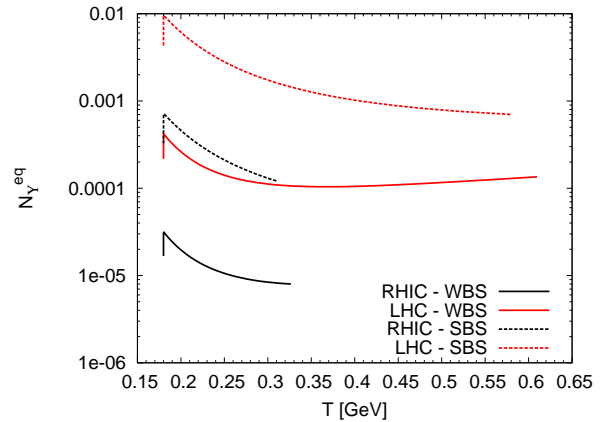}
   \caption{Equilibrium numbers of $\Upsilon$ vs.~temperature at RHIC
 ($\sqrt{s_{NN}}=0.2$\,TeV, lower two lines) and LHC 
($\sqrt{s_{NN}}=2.76\,$TeV, upper two lines) using equation \eqref{Neq} with
bottom-production cross sections as specified in Sec.~\ref{ssec_xsec} below. 
The solid lines correspond to the weak-binding scenario (with in medium 
$b$-quark mass) and the dashed lines to the strong-binding scenario (with 
vacuum $b$-quark mass).}
   \label{fig:Neq}
 \end{figure}

%%%%%%%%%%%%%%%%%%%%%%%%%%%%%%
To simulate the evolution of the bottomonium abundances in URHICs at
RHIC and LHC we adopt the rate-equation approach of 
Ref.~\cite{Grandchamp:2003uw} (also used in 
Ref.~\cite{Grandchamp:2005yw}). It describes the approach of the
number, $N_Y$, of bottomonium state $Y$, toward equilibrium as
\begin{equation}
\label{rate-eq}
\frac{\mathrm{d}N_{Y}}{\mathrm{d}\tau}=-\Gamma_{Y}^{\rm diss}(T)
\left[N_{Y}-N_{Y}^{\rm eq}(T)\right] \ .
\end{equation}
The key quantities in this equation are the inelastic reaction rate,
$\Gamma_{Y}^{\rm diss}$, and the equilibrium limit, $N_Y^{\rm eq}$, which
directly relate to the spectral properties discussed in the previous
section. The inelastic reaction rate would be all that is needed if
only suppression effects were considered, governed by the first (loss) 
term on the right-hand-side of Eq.~\eqref{rate-eq}. Regeneration 
processes are accounted though the gain term, 
$\Gamma_{Y}^{\rm diss}(T)N_{Y}^{\rm eq}(T)$. Its form is dictated by the 
detailed balance between the dissociation and regeneration processes. 
Apart from the inelastic reaction rate, the gain term is controlled
by the equilibrium limit which is computed as follows. For a given
collision energy, system and centrality, the total number of $b\bar b$
pairs is assumed to be constant throughout the evolution of the expanding 
fireball. Assuming relative chemical equilibrium between all available
states containing bottom quarks at given temperature and volume of the 
system, the $b\bar b$ number is matched to the equilibrium numbers
of bottom states by the condition 
\begin{equation}
\label{Nbb}
N_{b\bar b}=\frac{1}{2}N_{\rm op}\frac{I_{1}(N_{\rm op})}{I_{0}
(N_{\rm op})}+N_{\rm hid} \ ,  
\end{equation}
by means of a $b$-quark fugacity factor, 
$\gamma_{b}\equiv\gamma_{\bar b}$. The equilibrium open- and 
hidden-bottom numbers in the QGP are given by
\begin{eqnarray}
N_{\rm op}&=&\gamma_{b} \ V_{\rm FB} \ d_{b} 
\int\frac{d^3p}{(2\pi)^3} f^b(p;T) 
\\
N_{\rm hid}&=&\gamma_{b}^{2} \ V_{\rm FB}\sum\limits_Y d_Y \int 
\frac{d^3p}{(2\pi)^3} f^Y(p;T) \ , 
\label{Neq}
\end{eqnarray}
and $V_{\rm FB}$ is the fireball volume at given temperature, $T(\tau)$. 
The equilibrium number densities are simply those of $b$ quarks (with 
spin-color and particle-antiparticle degeneracy $d_b$=6$\times$2) and 
bottomonium states (summed over 
including their spin degeneracies, $d_Y$). % It is precisely here
% where the spectral properties of the $Y$ states figure via their
% real part, in particular the $b$-quark mass in $N_{\rm op}$. 
They are calculated using the same in-medium masses for open- and 
hidden-bottom states as in the spectral function analysis in 
Sec.~\ref{sec:sf} for both the SBS and WBS, respectively.
At fixed $N_{b\bar b}$ a large $b$-quark mass leads to
a large $\gamma_b$ which in turn results in a large equilibrium number, 
$N_Y$. 
This is reflected in the temperature dependence of the $\Upsilon$ 
equilibrium numbers which are much larger in the strong- than in 
the weak-binding scenario, cf.~Fig.~\ref{fig:Neq}.
%The softening of this difference for increasing temperature^M
%is a result of flatter thermal distribution functions in the weak binding scenario ^M
%as $m_{b\bar b}\to \frac{1}{2}m_{\Upsilon}$ ($\epsilon_B\to 0$). The equillibrium number^M
%increases towards $T_c$ as it becomes more favorable to have bound $b\bar b$ pairs than ^M
%free b and $\bar b$ quarks due to a decrease in $N_{op}$ and $\gamma_b$. The vertical line^M
%at $T_c$ is due to the decrease of $\gamma_b$ during the fixed temperature volume expansion ^M%of the mixed phase QGP.
The ratio of modified Bessel functions, $I_{1}(N_{op})/I_{0}(N_{op})$,
in Eq.~\eqref{Nbb} enforces exact conservation of the net $b$ 
number in each event (canonical ensemble). 
%For the number of hidden bottom states, $N_{hid}$,^M
%we include all bottomonium states listed in Ref.~\cite{Eidelman04}

In heavy-ion collisions, however, $b$-quarks are not expected to 
{\it kinetically} equilibrate~\cite{Rapp:2009my}. Bottom-quark spectra
which are harder than in equilibrium imply less phase-space overlap 
between  $b\bar b$ pairs and thus a reduced $Y$ regeneration. To account
for this we follow Ref.~\cite{Grandchamp:2002wp} by introducing
a relaxation-time factor
\begin{equation}
\label{eq:rfactor}
\mathcal{R}(\tau)=1-\exp\left(1-\int\limits_{\tau_0}^{\tau}
\frac{\mathrm{d}\tau'}{\tau_{\rm eq}}\right)
\end{equation}
whose timescale is set by the kinetic relaxation time, 
$\tau_{\rm eq}$, of $b$ quarks in the QGP. In our calculations it is 
taken as $\tau_{\rm eq}\simeq10$\,fm/$c$ around $T\simeq2\,T_c$, 
%to $\tau_{eq}\approxeq 1.1~fm/c$$1.0 GeV$ respectively, 
in a range consistent with microscopic $T$-matrix calculations
of bottom diffusion~\cite{Riek:2010fk} (with these diffusion coefficients
a fair description of the observed open heavy-flavor suppression and 
elliptic flow at RHIC is possible~\cite{vanHees:2007me}). 

%%%%%%%%%%%%%%%%%%%%%%%%%
\subsection{Fireball Model}
\label{ssec_fb}
%%%%%%%%%%%%%%%%%%%%%%%%%
Our fireball model has been updated compared to 
Ref.~\cite{Grandchamp:2005yw} by matching it to our recent charmonium 
applications at RHIC~\cite{Zhao:2010nk} and LHC~\cite{Zhao:2011cv}. 
We assume an isentropically expanding isotropic firecylinder whose 
eigenvolume is approximated by the ansatz 
\begin{equation}
\label{eq:vol-fb}
V_{\rm FB}(\tau)=\left(z_{0}+v_{z}\tau\right)\pi
\left(r_0+\frac{1}{2}a_{\perp}\tau^{2}\right)^{2} \ , 
\end{equation}
with initial transverse radius, $r_0$, longitudinal velocity 
$v_{z}$=1.4$c$  and transverse acceleration
$a_{\perp}$=0.1$c^2$/fm. The latter is larger than in 
Ref.~\cite{Grandchamp:2005yw} to better reproduce measured
bulk-hadron $p_t$ spectra and leads to a slight reduction in the QGP 
and mixed-phase lifetime.  At each centrality and collision energy 
of a heavy-ion collision, the total entropy, $S$, is chosen to 
reproduce the observed charged-hadron multiplicity (computed with a 
hadron resonance gas at a chemical freezeout temperature, 
$T_{\rm ch}$=$T_c$=180\,MeV). With $S=10000(22000)$ for 
$N_{\rm part}$=375 nucleon participants at RHIC (LHC) one obtains 
$dN_{\rm ch}/dy$=800(1750). 
The entropy density, 
$s(\tau)=S/V_{\rm FB}(\tau)$, is used to determine the temperature,
$T(\tau)$, by means of a QGP quasiparticle equation of state for $s(T)$. 
At $T$=$T_c$ a standard mixed-phase construction is employed. We have
verified that the impact of the subsequent hadronic evolution (also
in the mixed phase) is negligible for all $Y$ states. The 
initial longitudinal length, $z_0=\tau_0\Delta y$, controls the initial 
temperature ($\Delta y=1.8$ is the rapidity coverage of the fireball). 
With the QGP formation time $\tau_0=0.6(0.2)$\,fm/$c$ one obtains 
$T_0=330(610)$\,MeV at RHIC (LHC). 
\begin{figure}[!t]
   \centering
   \includegraphics[width=0.95\linewidth,clip=]{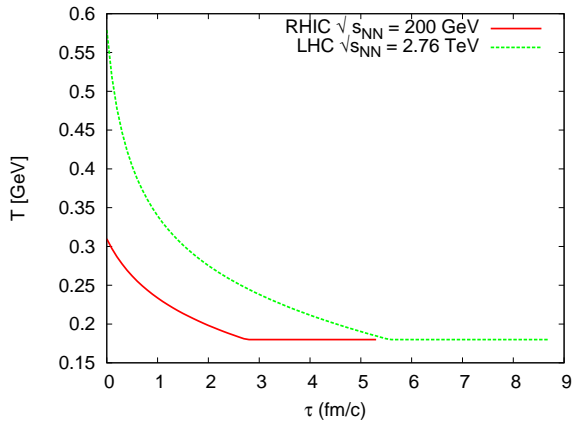}
   \caption{Temperature evolution of the fireball model at RHIC 
 (solid line) and LHC (dotted line). At RHIC the evolution is given for 
 central Au-Au collisions at $\sqrt{s_{NN}}$=0.2\,TeV and a QGP formation 
 time $\tau_0=0.6$~fm/$c$; at LHC the evolution is for central 
 Pb-Pb collisions at $\sqrt{s_{NN}}$=2.76\,TeV and $\tau_0=0.20~$fm/$c$.}
   \label{fig:temp-evol}
 \end{figure}
The temperature evolution of the fireball for central collisions is 
displayed in Fig.\ref{fig:temp-evol}, with QGP and mixed phases 
indicated. 
%However, in the case that the temperature dips below the 
%critical temperature, $T_c=180~MeV$, a mixed phase entropy density 
%is employed to retain correct evolution throughout the lifetime of 
%the QGP.

%%%%%%%%%%%%%%%%%%%%%%%%%%%%%%%%%%%%%%%%%%%%%
\subsection{Primordial $Y$ and Bottom Yields}
\label{ssec_xsec}
%%%%%%%%%%%%%%%%%%%%%%%%%%%%%%%%%%%%%%%%%%%%
The rate equation requires an initial condition for the number of
bottomonia at the QGP formation time, $N_Y(\tau_0)$. We determine these
by using their production cross section in $pp$ collisions and binary 
collision ($N_{\rm coll}(b)$) scaling in AA collisions. We use inelastic NN cross 
sections of $\sigma_{pp}^{\rm inel}$=42\,mb and 62\,mb at 
$\sqrt{s}$=0.2 and 2.76\,TeV, respectively. In addition, we allow for 
cold-nuclear-matter (CNM) suppression, collectively defined as being 
due to modifications of $N_Y$ prior to thermalization, most notably 
nuclear modifications of the parton distribution functions, absorption 
on passing-by primordial nucleons (or even secondary particles), and the 
Cronin effect. For simplicity, we approximate the combination of all 
CNM effects by a suppression factor, 
\begin{equation}
\label{snuc}
S_{\rm nuc} = \exp\left[-\rho_N\sigma_{\rm abs}L(b)\right]  \ ,
\end{equation}
with an effective nuclear absorption cross section, $\sigma_{\rm abs}$. 
The latter will be considered over a range of 1-3.1\,mb at RHIC and 
0-2\,mb at LHC. The former is compatible with the STAR measurement of
$R_{d\rm Au}=0.78\pm0.20\pm0.28$ for $\Upsilon$(1S+2S+3S) at 
RHIC~\cite{Liu:2009wa}, while
CNM effects are expected to be reduced at LHC due to the much increased 
Lorentz contraction of the incoming nuclei. 
The other parameters in Eq.~\eqref{snuc} are the nuclear density, 
$\rho_N$=0.14~fm$^{-3}$, and the impact-parameter dependent path length, 
$L(b)$, evaluated with a Glauber model for the nuclear overlap.

\begin{table}
 \begin{tabular*}{\linewidth}{ c | c | c | c | c}
        \hline
        \hline
 Quantity & $\sqrt{s_{NN}}$=0.2\,~TeV & Ref. & $\sqrt{s_{NN}}$=2.76\,TeV & Ref. 
\\  \hline
$\sigma_{pp}^{\rm inel}$ [mb] & 42 &\cite{Eidelman04} & 62 & \cite{Eidelman04} \\
$\sigma_{pp\to \Upsilon}^{\rm inc}$ [nb] & 6.6 & \cite{FUV08}* & 
$250\pm30 $ & \cite{Khachatryan11}* 
\\ 
$\sigma_{pp\to \Upsilon}^{\rm dir}$ [nb] & 3.4 & \cite{FUV08}* & 128 & \cite{Vogt10}* 
\\
$\sigma_{pp\to \Upsilon^{\prime}}^{\rm dir}$ [nb] & 2.2 & \cite{FUV08}* & 
83 & \cite{Vogt10}* 
\\
$\sigma_{pp\to \chi_b}^{\rm dir}$ [nb] & 7.1 & \cite{FUV08}* & 270 & 
\cite{Vogt10}* 
\\
$\sigma_{pp\to b\bar b}^{\rm tot}$ [$\mu$b] & $3.2\pm1.8$ 
& \cite{Adare:2009ic}* & $142\pm 25$ & \cite{LHCb10}* 
\\
$R_y$ & 0.52 & \cite{Vogt03} & 0.3 & \cite{Vogt03} 
\\
$S$($N_p$=375) & 10000 & \cite{Grandchamp:2001pf} & 22000 & \cite{Zhao:2011cv}  
\\ 
$\tau_0$ [fm/$c$] & 0.6 & \cite{Zhao:2010nk} & 0.2 & \cite{Zhao:2011cv}  
\\
        \hline
        \hline
        \end{tabular*}  
\caption{Summary of cross sections in $pp$ collisions used to 
initialize the hidden- and open-bottom abundances of our rate equations 
in Au-Au at RHIC (second column) and Pb-Pb at LHC (fourth column). 
Uncertainties are quoted when quantitatively available; starred 
references require further explanation, given in the text.}   
\label{tab:inputs}
\end{table}

Current experimental limitations prevent resolving of individual 
$\Upsilon$($n$S) states at RHIC and are reported as a combined cross 
section for $\Upsilon$(1S+2S+3S) including the 
$\Upsilon(n\mathrm{S})\to e^+e^-$ branching ratios, i.e., 
$\sum_{n=1}^3 \mathcal{B}(n\mathrm{S})
\sigma(n\mathrm{S})$~\cite{Abelev:2010am}.
The measured value turns out to be consistent with next-to-leading 
order (NLO) color-evaporation model (CEM) calculations (using MRST HO 
parton distribution functions) for the individual $\Upsilon$ production 
cross sections~\cite{FUV08}. We therefore adopt the latter in our 
calculations, which are close to the values used in previous 
work~\cite{Grandchamp:2005yw}.

To estimate the bottomonium cross sections at $\sqrt{s_{NN}}$=2.76\,TeV 
we start from the inclusive production of $\Upsilon$ measured by 
CMS~\cite{Khachatryan11}, 
$\sigma_{pp\to \Upsilon}^{\rm tot}$($\sqrt{s}$=7\,TeV,$|y|$$<$2)\,$\simeq$\,(300$\pm$40)\,nb
(employing a $\mu^+\mu^-$ branching ratio of $B_{\mu\mu}$=2.5\%).
Guided by NLO calculations~\cite{Vogt09} we extrapolate
to 2.76\,TeV by a factor of 0.5 and to full rapidity by a factor 
of 1/0.6. For a rapidity window of $\Delta y$=1.8 for one thermal
fireball ($R_y$=0.3, see below), we obtain
$\sigma_{pp\to\Upsilon}^{\rm tot}$($\sqrt{s}$=2.76\,TeV,$|y|$$<$0.9)\,$\simeq$\,(75$\pm$8)\,nb. 
Alternatively, we can estimate this quantity by taking the CMS
midrapidity value of $d\sigma_{pp\to \Upsilon}/dy$($\sqrt{s}$=7\,TeV)\,$\simeq$\,(80$\pm$9)\,nb, 
downscale it by a factor of 2 and multiply by $\Delta y$=1.8 to obtain 
$\sigma_{pp\to \Upsilon}^{\rm tot}$($\sqrt{s}$=2.76\,TeV,$|y|$$<$0.9)\,$\simeq$\,(72$\pm$8)\,nb, 
consistent with the above estimate. These estimates are also compatible with the CDF 
value~\cite{Acosta:2001gv} of 
$d\sigma_{pp\to \Upsilon}/dy$($\sqrt{s}$=1.8\,TeV)\,$\simeq$\,(27$\pm$1.5)\,nb, 
extrapolated to 2.76\,TeV with a factor of $\sim$4/3.   
%The same extrapolation was used for 
%the $\sqrt{s_{NN}}$=7~TeV $b\bar b$ production cross section in 
%Ref.~\cite{LHCb10}. 
Based on the inclusive $\Upsilon$ production, we fix the cross sections 
for direct $\Upsilon$, $\Upsilon^{\prime}$, and $\chi_b$(1P) to 
maintain compatibility with the feeddown fractions given in 
Ref.~\cite{Affolder00} and listed in Tab.~\ref{table:feeddown};
e.g., with a ca.~30\% branching for $\Upsilon^{\prime}\to\Upsilon X$ 
one needs $\sigma_{pp\to \Upsilon'} \simeq \frac{1}{3} 
\sigma_{pp\to \Upsilon}$ to obtain a feeddown fraction of $\sim$10\%,
and likewise for the $\chi_b$ states.

The regeneration component requires the knowledge of the $b\bar b$ 
cross section, which we also take from $pp$ measurements together with 
binary-collision scaling. 
At RHIC the STAR measurement~\cite{Xu11} gives
$\sigma_{pp\to b\bar b}^{\rm tot}$($\sqrt{s}$=0.2\,TeV)=(1.6$\pm$0.5)\,$\mu$b,
where we combined statistical and systematic errors. This is slightly 
smaller than, but compatible with, the value of 2\,$\mu$b used in 
Ref.~\cite{Grandchamp:2005yw}. The recent PHENIX measurements 
finds a larger central value~\cite{Adare:2009ic}, 
$\sigma_{pp\to b\bar b}^{\rm tot}$($\sqrt{s}$=0.2\,TeV)=(3.2$\pm$1.8)\,$\mu$b,
which, however, is still compatible with the STAR datum within combined 
statistical and systematic errors~\footnote{The PHENIX measurement of the
total $b\bar b$ cross section is based on the midrapidity datum of
$d\sigma_{pp}^{b\bar b}/dy$=(0.92$\pm$0.5)\,$\mu$b; the extrapolation to 
the total is consistent with our rapidity-coverage factor of 
$\Delta y/R_y$=1.8/0.52=3.46, see Eq.~\eqref{eq:n-prod}, as originally 
inferred in Ref.~\cite{Grandchamp:2005yw}}. In our calculations below
we will use a PHENIX value since it results in a $\Upsilon/b\bar b$ 
ratio which is closer to the LHC value, but we will also comment on
using the STAR value.
At LHC our baseline value is obtained from LHCb data at 
7\,TeV~\cite{Khachatryan11}, again downscaled by a factor of 2.
With $R_y$=0.3 for one thermal fireball (see below), one then finds
$\sigma_{pp\to b\bar b}$($\sqrt{s}$=2.76\,TeV,$|y|$$<$0.9)$\simeq$(43$\pm$8)\,$\mu$b. 
Alternatively, one can estimate this quantity by using the midrapidity 
values of (35$\pm$5)\,$\mu$b at 7\,TeV~\cite{Khachatryan11} and 
(15$\pm$2)\,$\mu$b at 1.96\,TeV~\cite{Acosta:2004yw} to extrapolate to 
$\sim$(20$\pm$3)\,$\mu$b at 2.76\,TeV (per unit rapidity), which, upon 
multiplying with $\Delta y$=1.8, yields 
$\sigma_{pp\to b\bar b}$($\sqrt{s}$=2.76\,TeV,$|y|$$<$0.9)\,$\simeq$\,(36$\pm$5)\,$\mu$b, 
consistent with the above value.
Since nuclear shadowing is believed to be significant at LHC, and
applicable to the initial $b\bar b$ production, a reduction factor
of $0.75$ will be included in Pb-Pb collisions (note that for
$\Upsilon$ states the shadowing effect is associated with
the CNM suppression factor, $S_{\rm nuc}$, in Eq.~\eqref{snuc}).
At RHIC, we assume no shadowing on initial $b\bar b$ production.

As indicated above, a thermal fireball of light particles covers about 
1.8 units in rapidity; the total number of initially produced $Y$ (or 
$b\bar b$) states within the  fireball are thus obtained as
\begin{equation}
\label{eq:n-prod}
N_X(\tau_0)=
\frac{\sigma_{pp\to X}^{\rm tot}}{\sigma_{pp}^{\rm inel}} \ N_{\rm coll}(b) 
\ R_y \ S_{\rm nuc}
\end{equation}
with $X=\Upsilon, \Upsilon^\prime, \chi_b, b\bar b$; here, 
$R_y\simeq0.52(0.3)$ denotes the fraction of the total number of $X$ 
contained in $\Delta y$=1.8 around midrapidity at RHIC (LHC). 
For $X=b\bar b$, $S_{\rm nuc}$=1(0.75) for RHIC (LHC).

To properly account for the observed inclusive $\Upsilon$ yields,
feeddown from higher bound states is computed. The magnitude of this 
contribution in $pp$ collisions is given in Tab.~\ref{table:feeddown}, 
reaching ca.~50\%  for the inclusive $\Upsilon$ yield. We assume this 
to carry over to the initial conditions for AA collisions and solve 
individual rate equations for direct $\Upsilon$,
$\Upsilon^{\prime}$ and $\chi_b$; we do not distinguish individual
$\chi_b$ states so that their combined feeddown amounts to 
37\% in $pp$; the $\Upsilon^{\prime}$ contributes 11\%. 
The feeddown from the $\Upsilon^{\prime\prime}$ is considered 
negligible (1\%). This decomposition of the total $\Upsilon$ yield is 
assumed to be independent of the collision energy and colliding systems.
\begin{table}[!t]
	\begin{tabular*}{0.9\linewidth}{@{\extracolsep{\fill}}  l | r}
	\hline
	\hline
	Prompt $\Upsilon$(1s) & $\sim 51\%$ \\
	$\Upsilon$(1s) from $\chi_b$(1P) decays & $\sim 27\%$ \\
	$\Upsilon$(1s) from $\chi_b$(2P) decays & $\sim 10\%$ \\
	$\Upsilon$(1s) from $\Upsilon$(2S) decays & $\sim 11\%$ \\
	$\Upsilon$(1s) from $\Upsilon$(3S) decays & $\sim 1\%$ \\
	\hline 
	\hline
	\end{tabular*}
\caption{The decomposition of the total (inclusive) $\Upsilon$(1s) yield into
the direct $\Upsilon$ and feeddown from excited states. The composition
was determined from $\sqrt{s_{NN}}$=39~GeV $p$-$p$
collisions~\cite{Affolder00}.}
\label{table:feeddown}
\end{table}

%%%%%%%%%%%%%%%%%%%%%%%%%%%%%%%%%%%%%%%% 
\section{Comparison to Data}
\label{sec:exp}
%%%%%%%%%%%%%%%%%%%%%%%%%%%%%%%%%%%%%%%%
Utilizing the previously discussed binding scenarios and dissociation 
mechanisms (Sec.~\ref{sec:sf}), implemented into the rate equation in 
a thermal-fireball background (Secs.~\ref{ssec_rate} and 
\ref{ssec_fb}), observables are calculated for comparison to recent 
data at RHIC and LHC in the form of the nuclear modification factor,
 \begin{equation}
\label{eq:raa}
R_{AA}(N_{\rm part})=\frac{N_Y^{AA}(N_{\rm part})}
{N_{\rm coll}(N_{\rm part}) \ \frac{\sigma_{pp\to Y}^{\rm tot}}
	{\sigma_{pp}^{inel}} \ R_y}  \ .
	\end{equation}
%This gives the total $\Upsilon$ state production normalized by  
%$\Upsilon$ production in $pp$ collisions, scaled by the number of 
%$NN$ collisions in the $A-A$ collision ($N_{coll}$) and modified by
%the $\Upsilon$ rapidity distribution, $R_y$. 
%If the production
%behaves naively in $A-A$ collisions, the scaled $pp$ collision should match
%observed yield exactly, giving a result of unity. 
The numerator, $N_Y^{AA}$, is the solution of the rate 
equation~(\ref{rate-eq}) at chemical freezeout. Deviations from unity 
are induced by CNM effects, suppression and 
regeneration in heavy-ion collisions. All previous studies of 
$R_{AA}^\Upsilon$ thus far show suppression to be the dominant effect;
significance of regeneration in the total yield was only suggested
in our previous calculation~\cite{Grandchamp:2005yw} and in the
statistical model~\cite{Andronic:2006ky} where all primordial bottomonia
are assumed to be dissociated.
In the following figures, our calculations for $R_{AA}$ are
broken up into nuclear absorption, primordial, and regeneration
components, a luxury not available to experiment, but instructive
for interpreting the data. 
%The nuclear modification
%factor is also examined for $\Upsilon^\prime$ and $\chi_b$(1p), as well
%as the $\Upsilon$(1s) state without feed-down effects (direct yield), 
%in order to quantify the source of the $\Upsilon$ suppression in both binding scenarios.

%%%%%%%%%%%%%%%%%%%%%%%%%%%
\subsection{RHIC}
\label{sec:rhic}
\begin{figure*}[!h]
\centering
\includegraphics[width=0.43\linewidth,clip=]{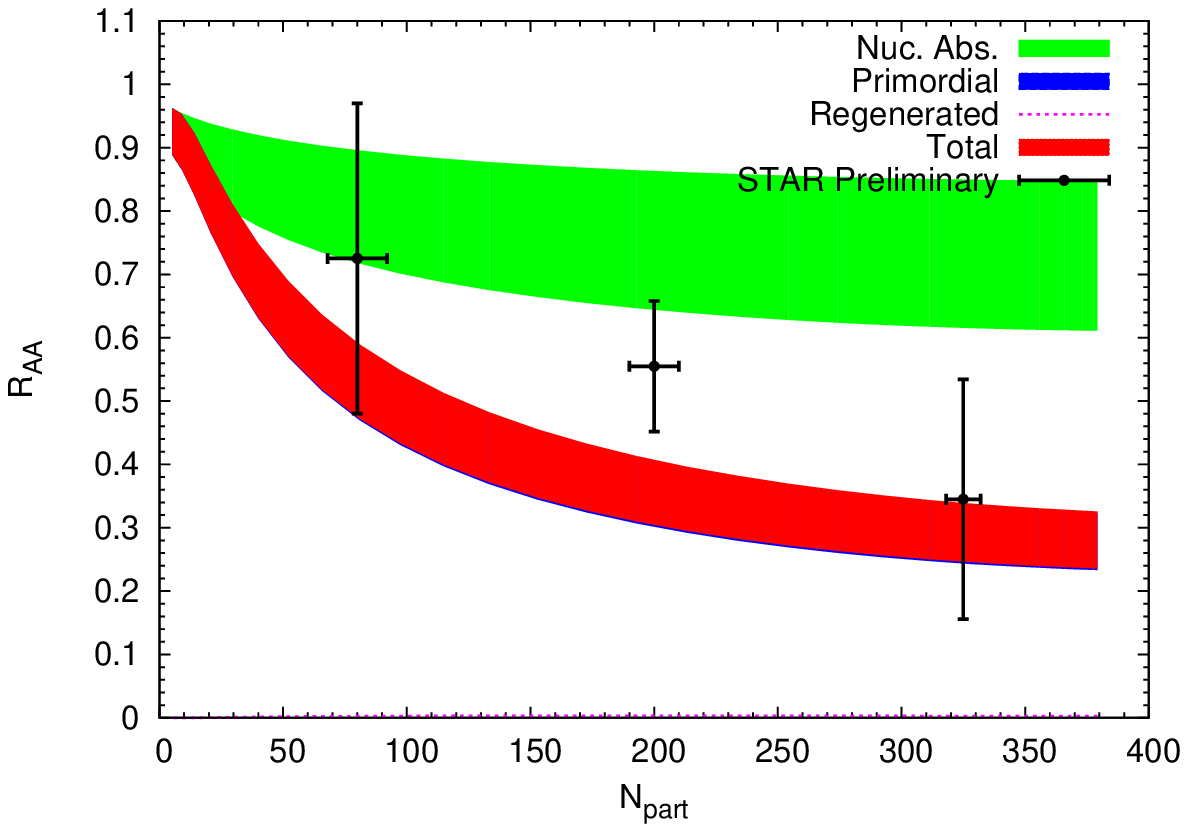}
\vspace{-0.2cm}
\includegraphics[width=0.43\linewidth,clip=]{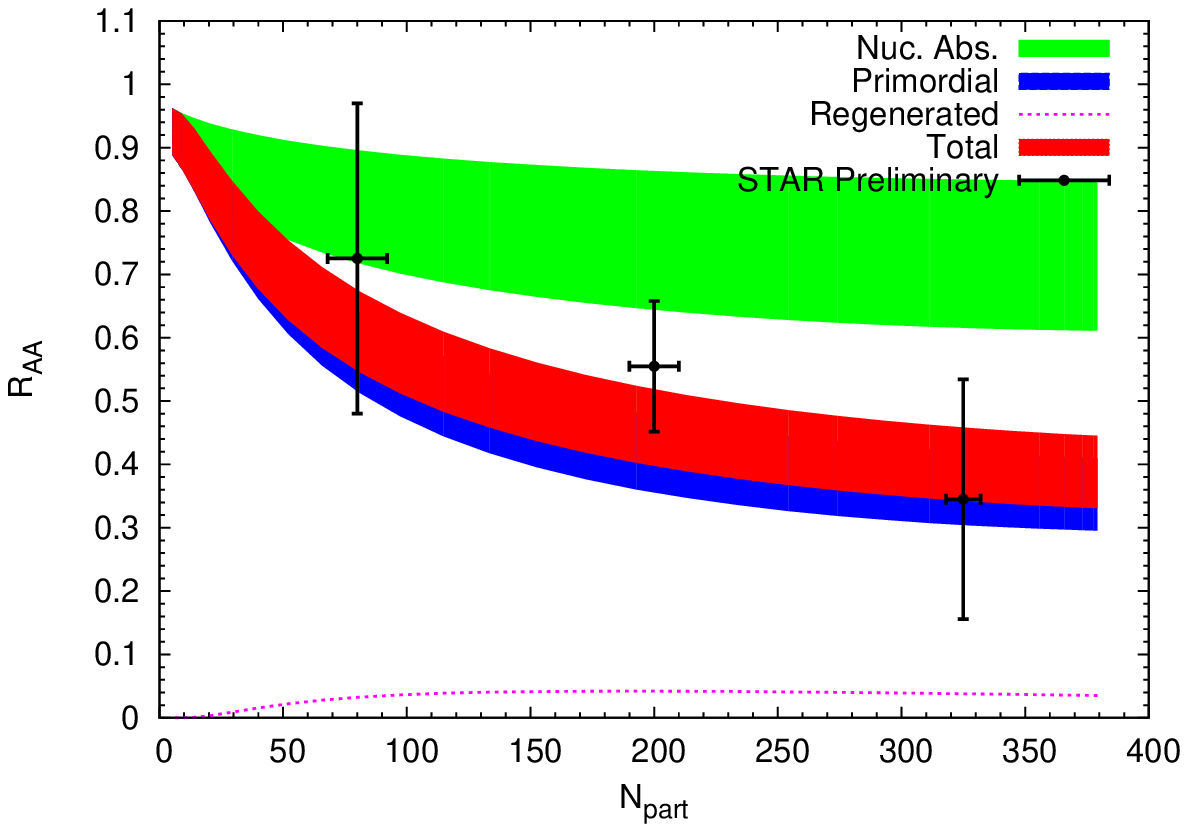}
\vspace{-0.1cm}
\includegraphics[width=0.43\linewidth,clip=]{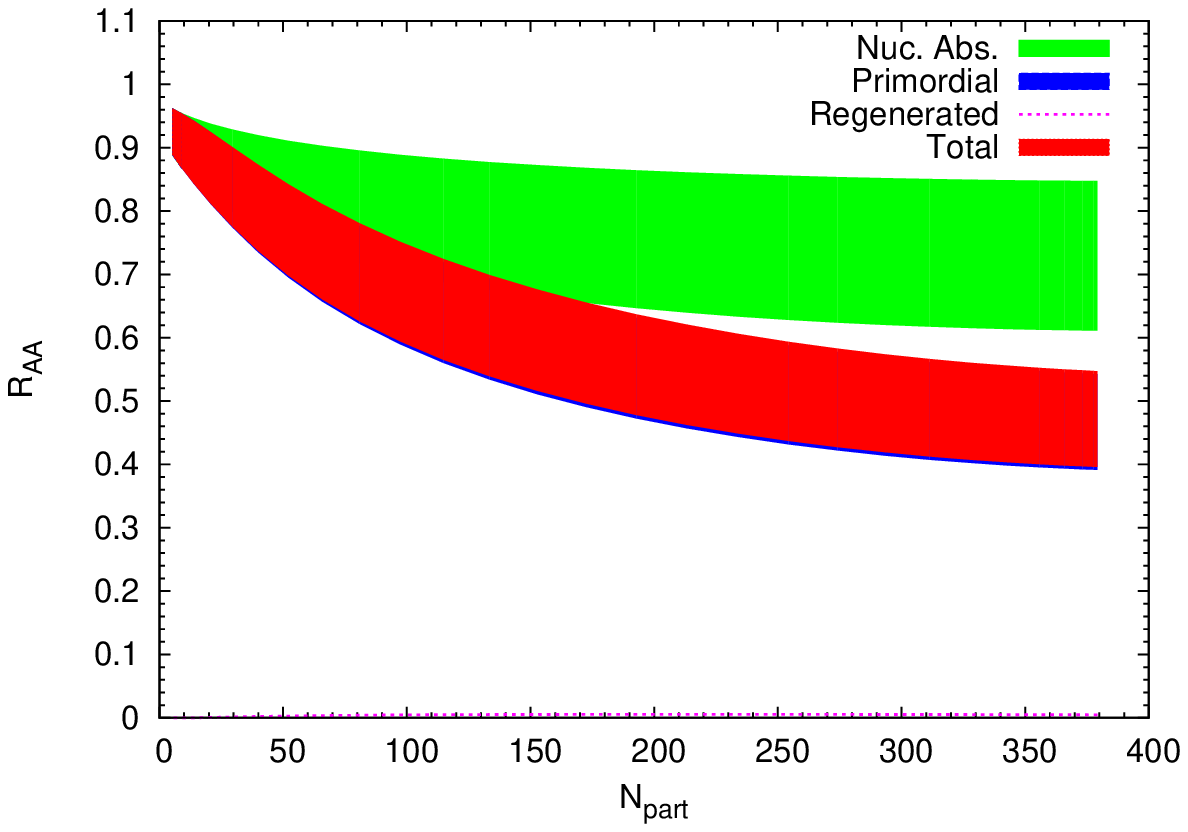}
\vspace{-0.1cm}
\includegraphics[width=0.43\linewidth,clip=]{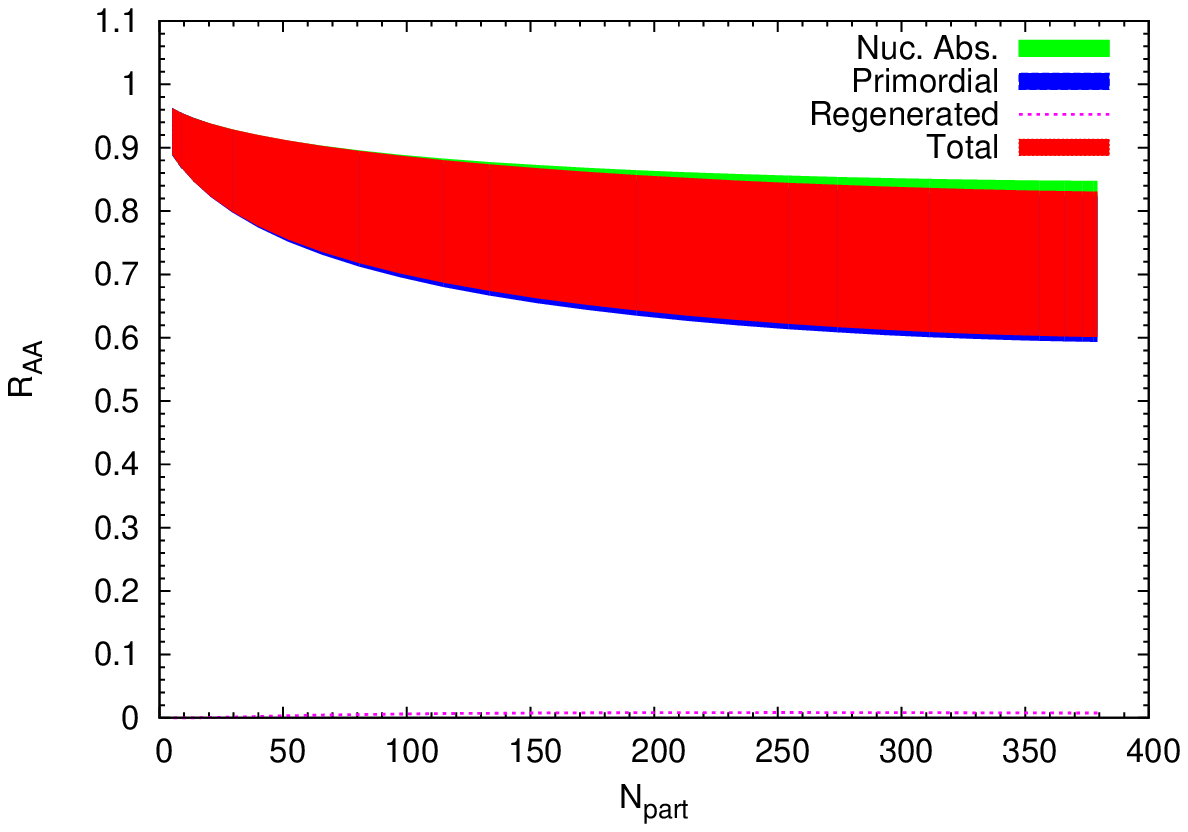}
\vspace{-0.1cm}
\includegraphics[width=0.43\linewidth,clip=]{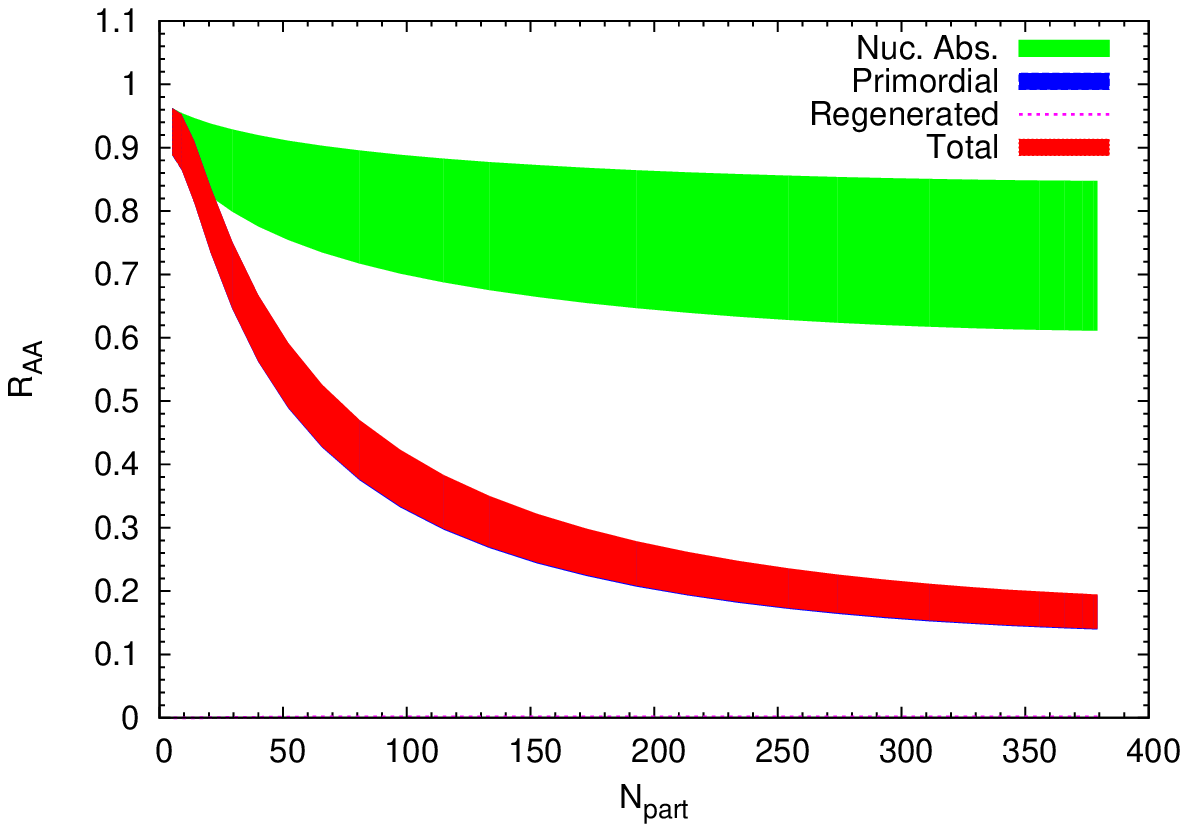}
\vspace{-0.1cm}
\includegraphics[width=0.43\linewidth,clip=]{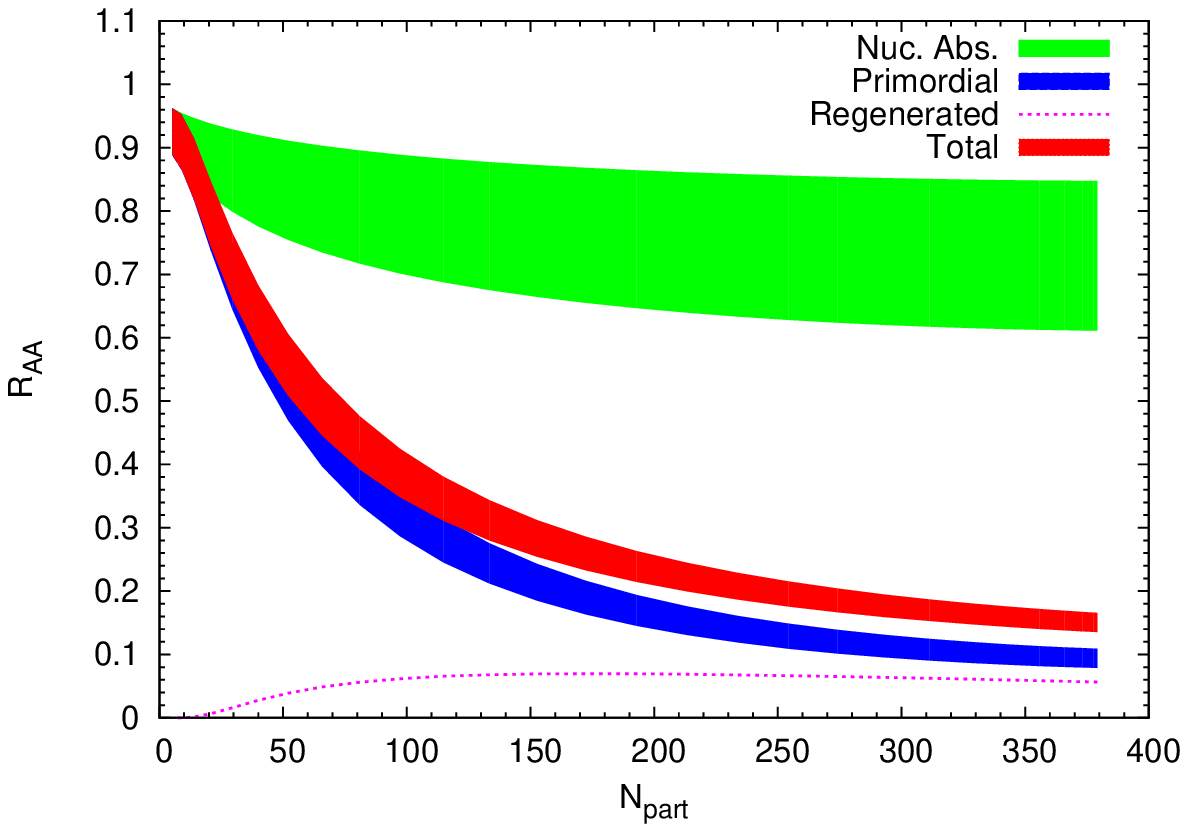}
\vspace{-0.1cm}
\includegraphics[width=0.43\linewidth,clip=]{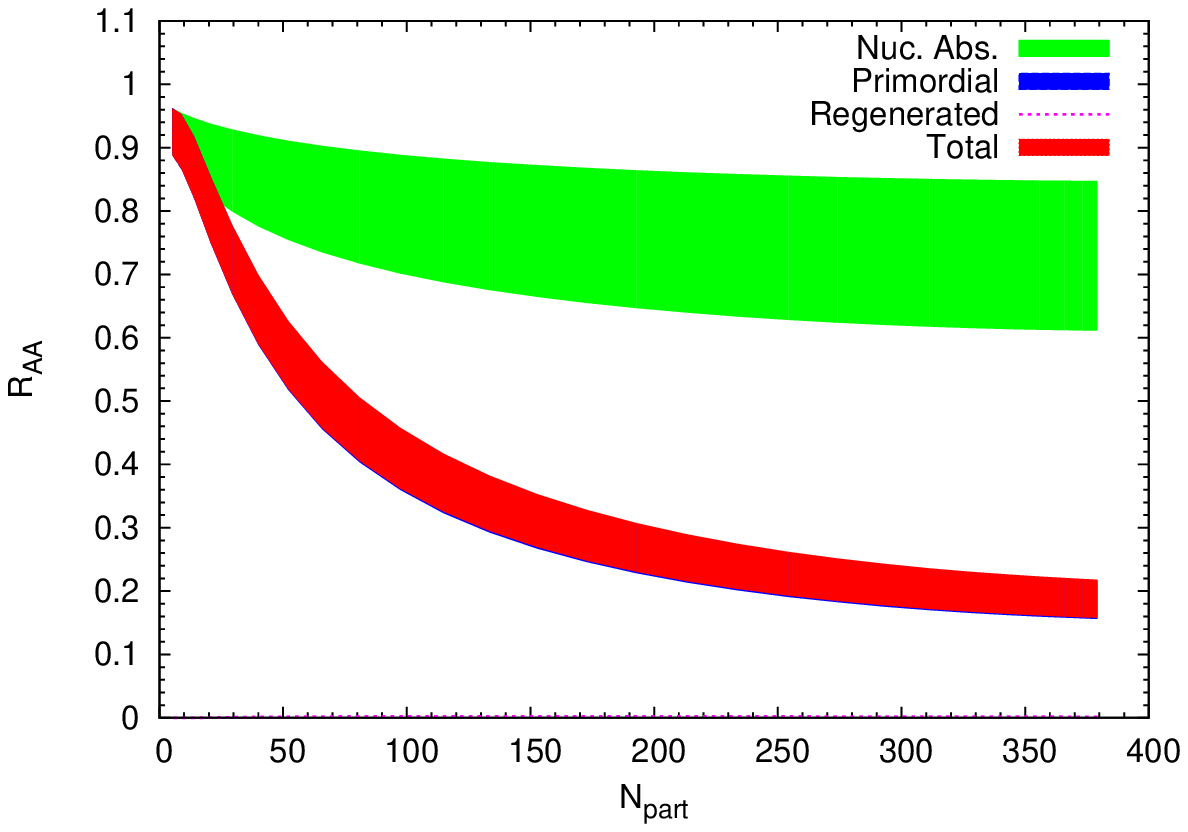}
\vspace{-0.0cm}
\includegraphics[width=0.43\linewidth,clip=]{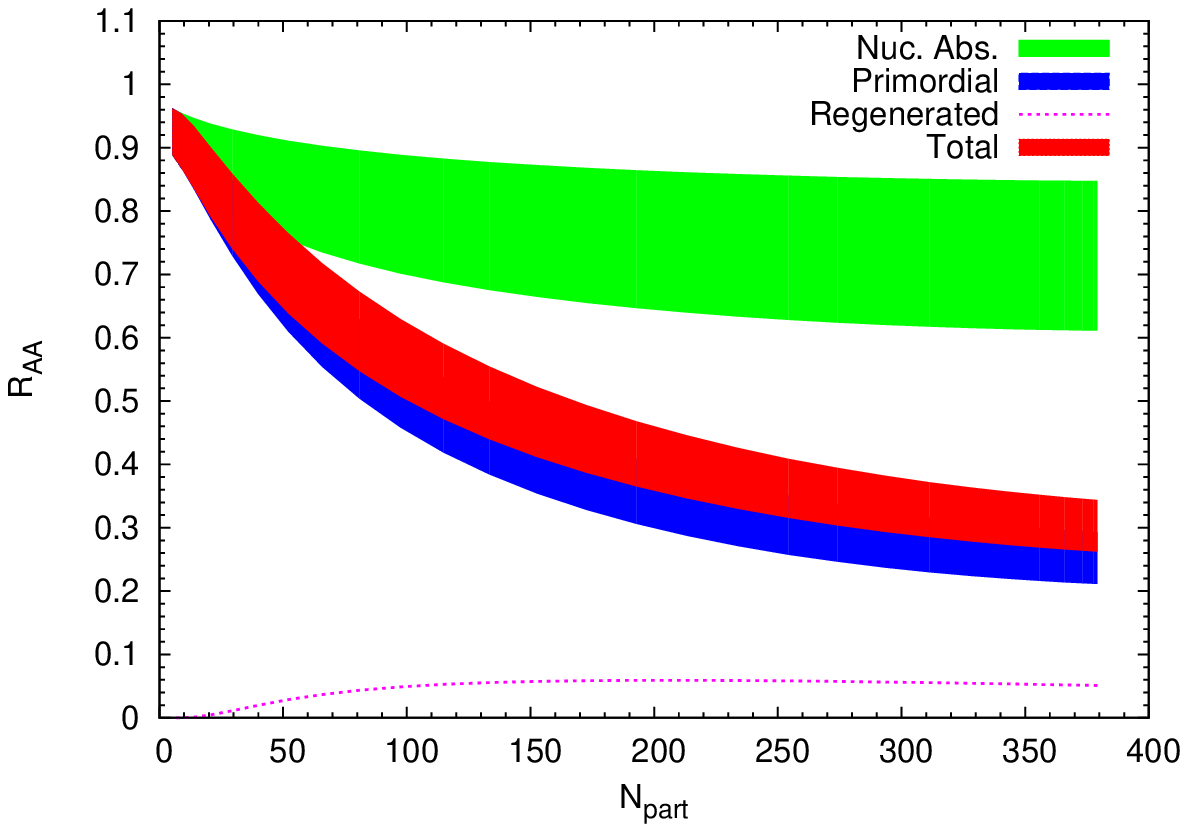}
\caption{The nuclear modification factor for $\Upsilon$(1S+2S+3S) including
$e^+e^-$ branching (top row, compared to STAR data~\cite{star-11}), direct 
$\Upsilon$ (second 
row), $\Upsilon^{\prime}$ (third row) and $\chi_b$ (bottom row), as a 
function of centrality in Au-Au($\sqrt{s_{NN}}$=0.2\,TeV) collisions at 
RHIC. The left column corresponds to the weak-binding scenario, the 
right one to the strong-binding scenario. In each panel, CNM effects 
alone are shown by the green band, CNM plus QGP suppression by the blue 
band, regeneration by the dashed pink line and the total by the red 
band. If regeneration is small, red and blue bands largely overlap.
} 
\label{fig:rhic-raa}
\end{figure*}

%%%%%%%%%%%%%%%%%%%%%%%%%%%
As mentioned above our calculations at RHIC supplement previous 
work~\cite{Grandchamp:2005yw} for the same collision system, with 
slightly updated fireball, spectral-function and input parameters. 
In Fig.~\ref{fig:rhic-raa} we summarize our results for the 
$R_{AA}(N_{\rm part})$ in 0.2~ATeV Au-Au collisions, for the combined 
$\Upsilon$(1S+2S+3S) (including $e^+e^-$ branching ratios)~\footnote{The
branching ratios are taken from Ref.~\cite{FUV08}; the suppression 
pattern of the 3S state is approximated as following the 2S one.} 
compared to STAR data in the upper row, and for the direct $\Upsilon$, 
$\Upsilon'$ and $\chi_b$ in rows 2-4. The left panels, representing the 
weak-binding scenario, show substantial suppression, reaching a factor 
of $\sim$2 for the ground state and more than 5 for the excited states 
in central Au-Au. Regeneration does not play any role due to a small 
equilibrium limit, recall Fig.~\ref{fig:Neq}. The combined $R_{AA}$(1S+2S+3S) 
is in the lower parts of, but not inconsistent with, the STAR data. 
In the strong-binding scenario the suppression of the ground state is 
noticeably less pronounced than in the WBS, being almost entirely due 
to CNM effects. On the other hand, it might seem surprising that the 
suppression of the excited states is rather comparable in both 
scenarios. As discussed in Sec.~\ref{sec:sf}, this is due to the similar 
dissociation rates of $Y(2S)$ and $\chi_b(1P)$ in SBS and WBS at 
moderate QGP temperatures (recall Fig.~\ref{fig:gamma}), originating 
from the larger coupling and favorable dissociation kinematics in the 
gluo-dissociation processes (thus compensating the larger binding 
energy). This, in particular, puts into question a ``sequential 
suppression pattern" (or ``thermometer") which is often discussed 
without taking into account the finite width of the bound states. 
The weaker suppression in the combined $R_{AA}$(1S+2S+3S) in the
SBS relative to the WBS is thus mainly due the $Y$(1S). This supports 
the assertion made in the original work~\cite{Grandchamp:2005yw}
that the suppression level of the ground-state $\Upsilon$ is 
the most sensitive observable for color screening. In the SBS, 
regeneration contributions are small but rather significant especially
for the $\chi_b$, helped by its large inelastic reaction rate. They 
decrease by a factor of 2 upon decreasing the bottom cross section 
from 3.2\,$\mu$b (PHENIX central value) to 1.6\,$\mu$b (STAR central 
value), since bottom states are in the canonical limit at RHIC. 

%%%%%%%%%%%%%%%%%%%%%%%%%%
\subsection{LHC Results}
\begin{figure*}[!h]
\centering
\includegraphics[width=0.43\linewidth,clip=]{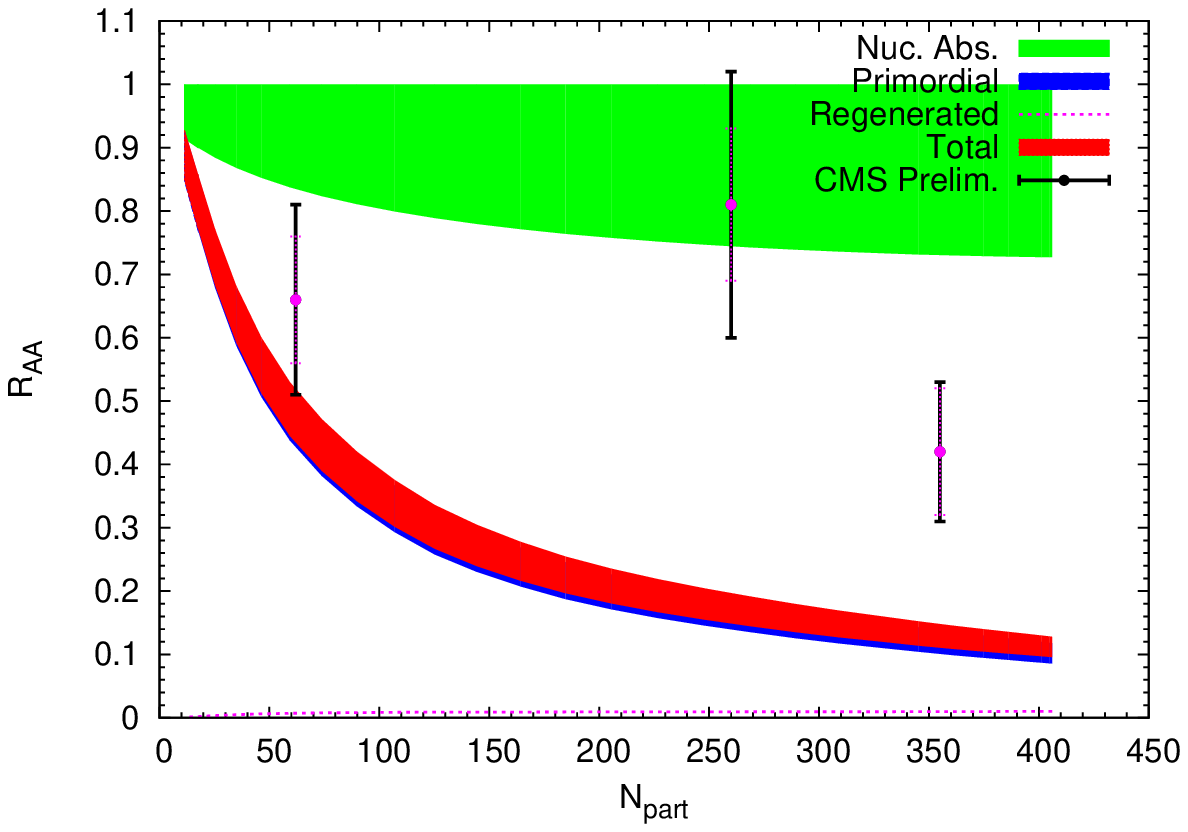}
\vspace{-0.2cm}
\includegraphics[width=0.43\linewidth,clip=]{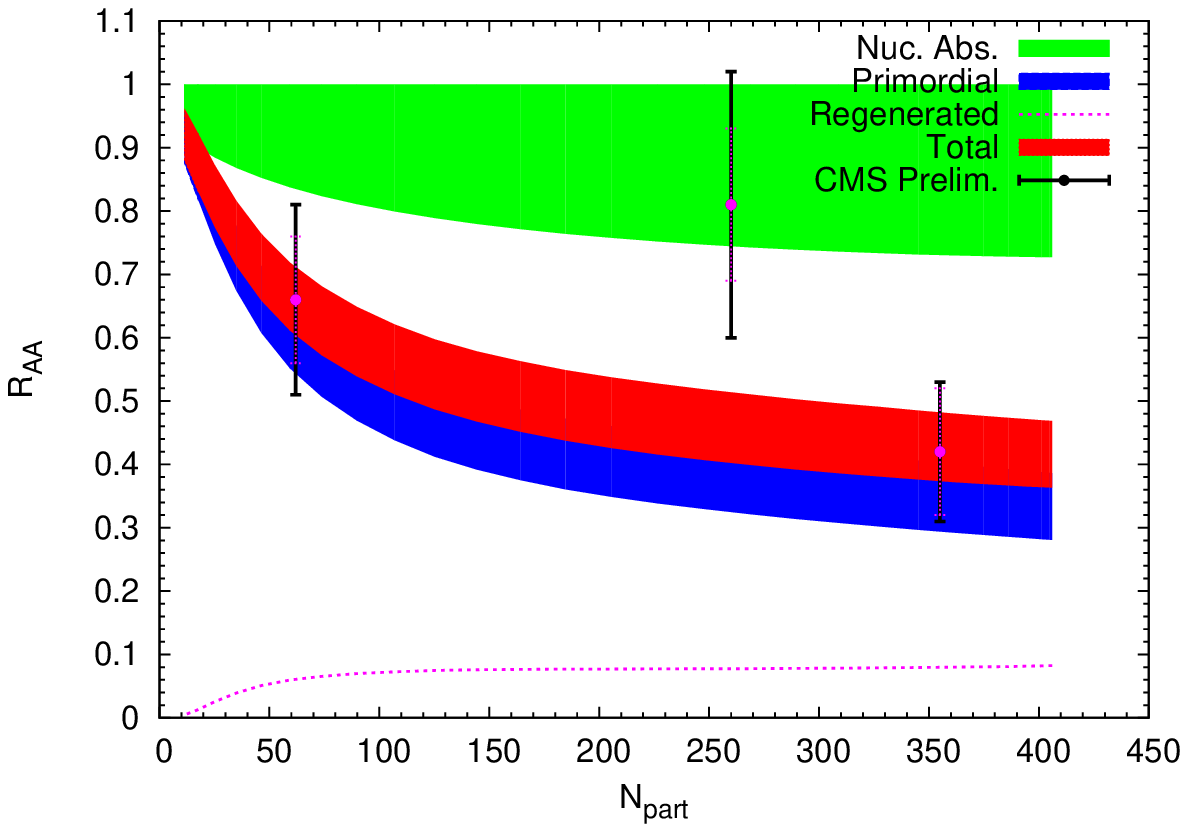}
\vspace{-0.1cm}
\includegraphics[width=0.43\linewidth,clip=]{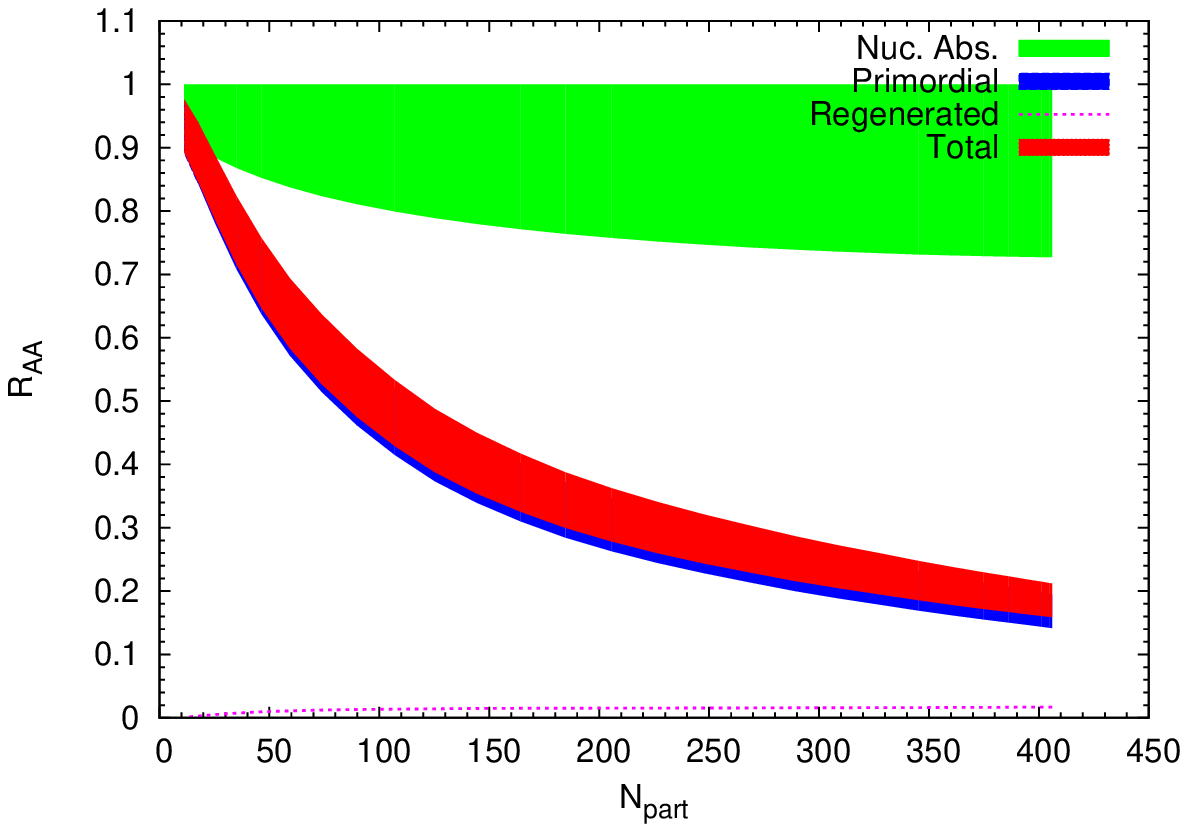}
\vspace{-0.1cm}
\includegraphics[width=0.43\linewidth,clip=]{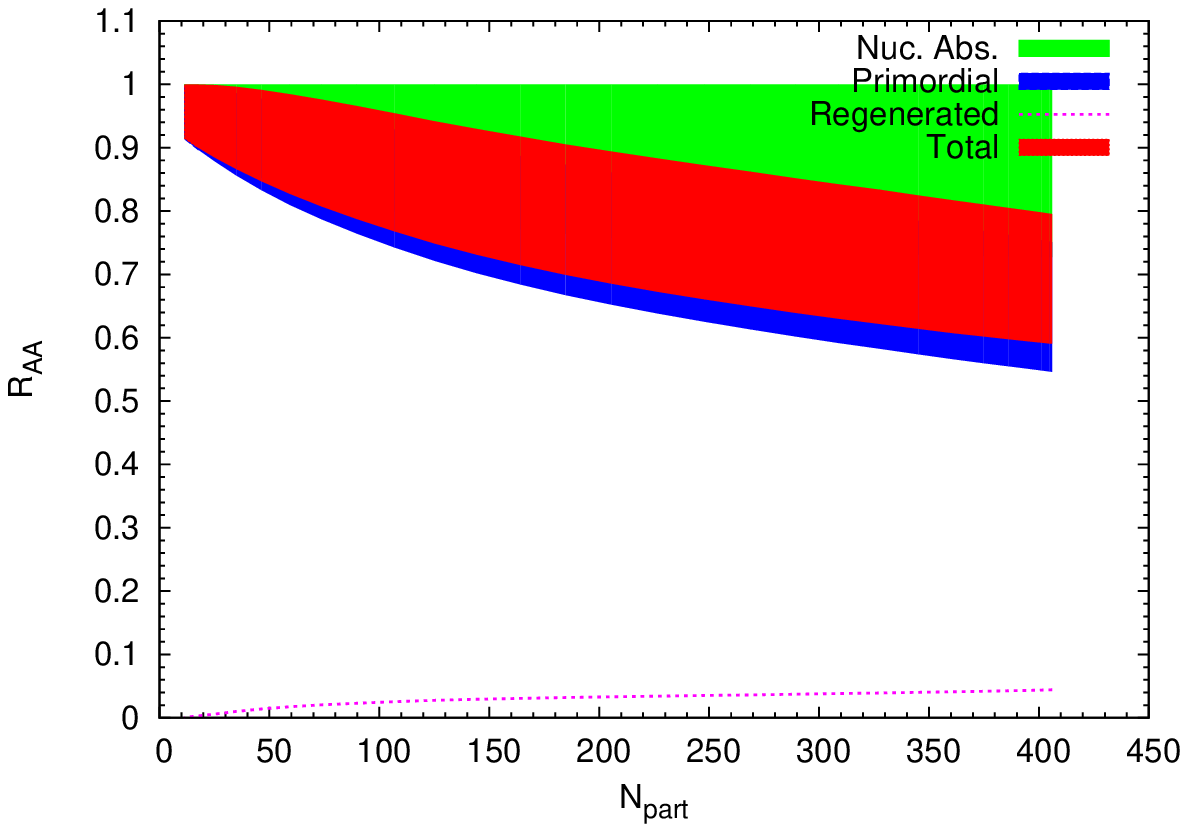}
\vspace{-0.1cm}
\includegraphics[width=0.43\linewidth,clip=]{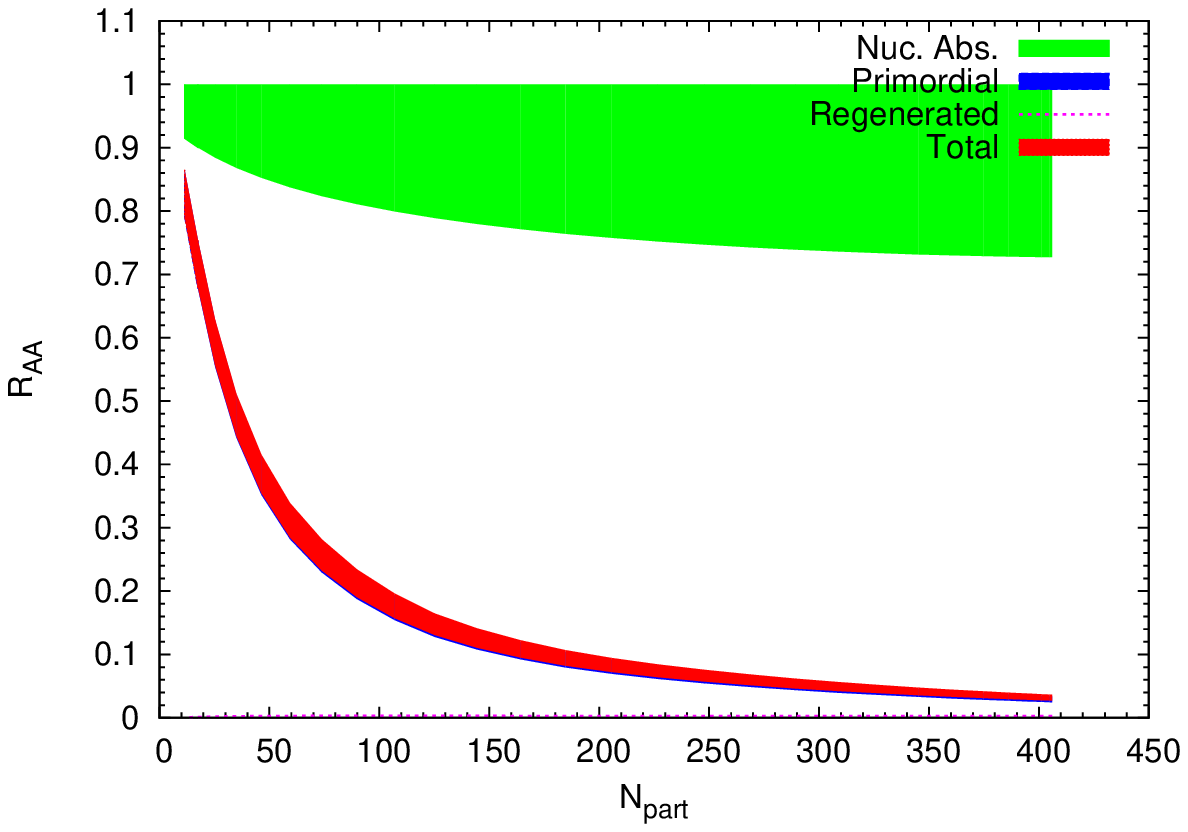}
\vspace{-0.1cm}
\includegraphics[width=0.43\linewidth,clip=]{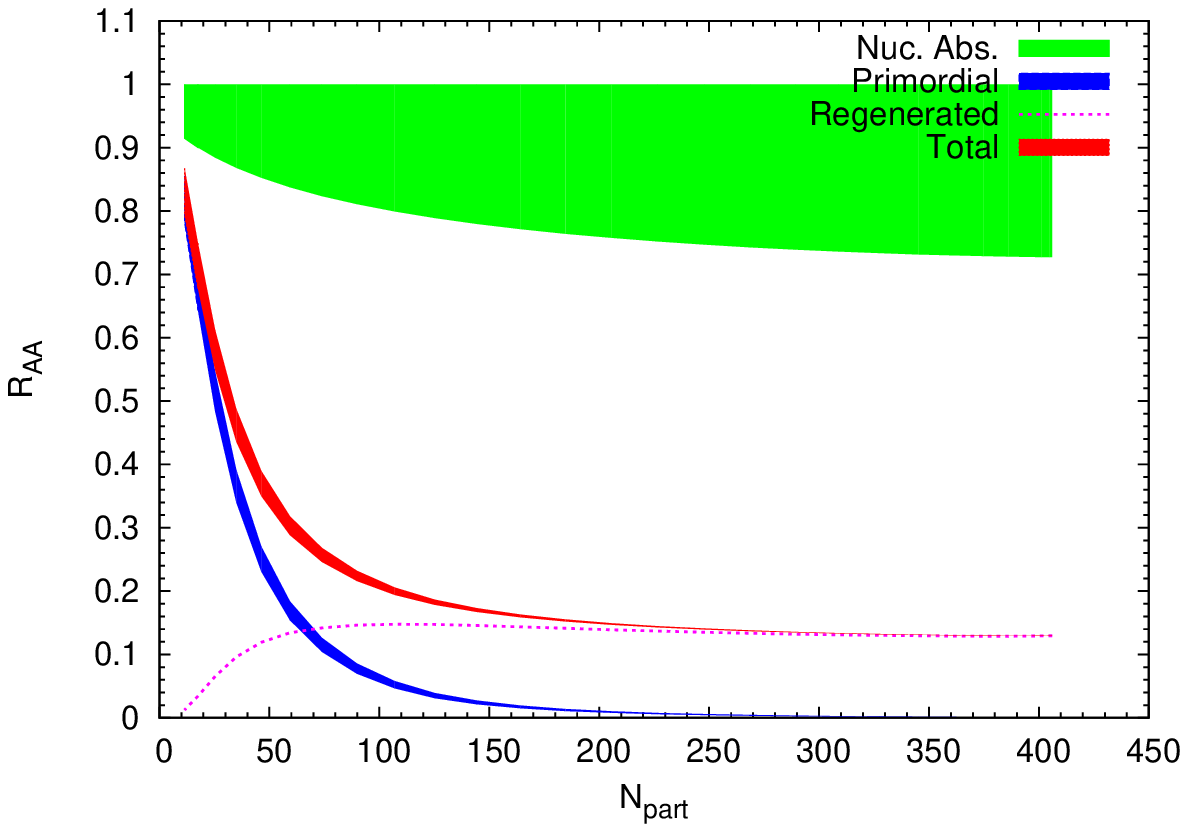}
\vspace{-0.1cm}
\includegraphics[width=0.43\linewidth,clip=]{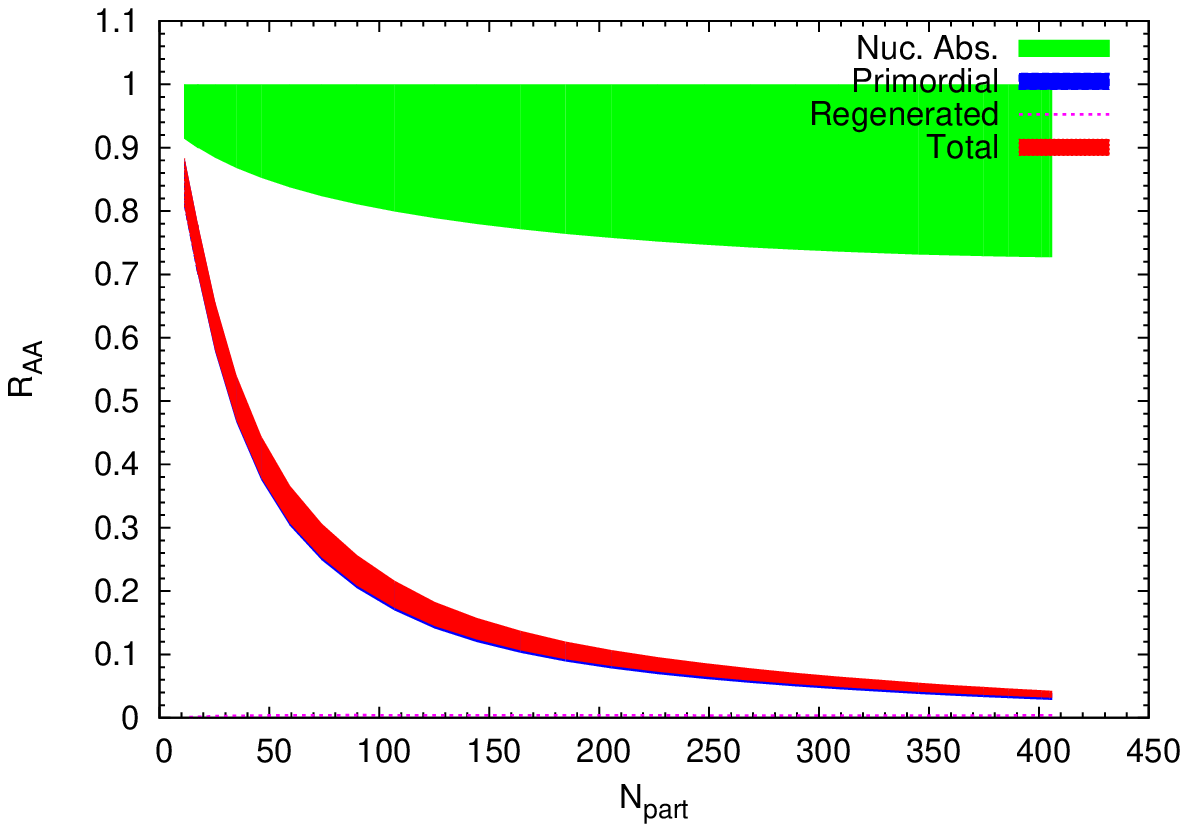}
\vspace{-0.1cm}
\includegraphics[width=0.43\linewidth,clip=]{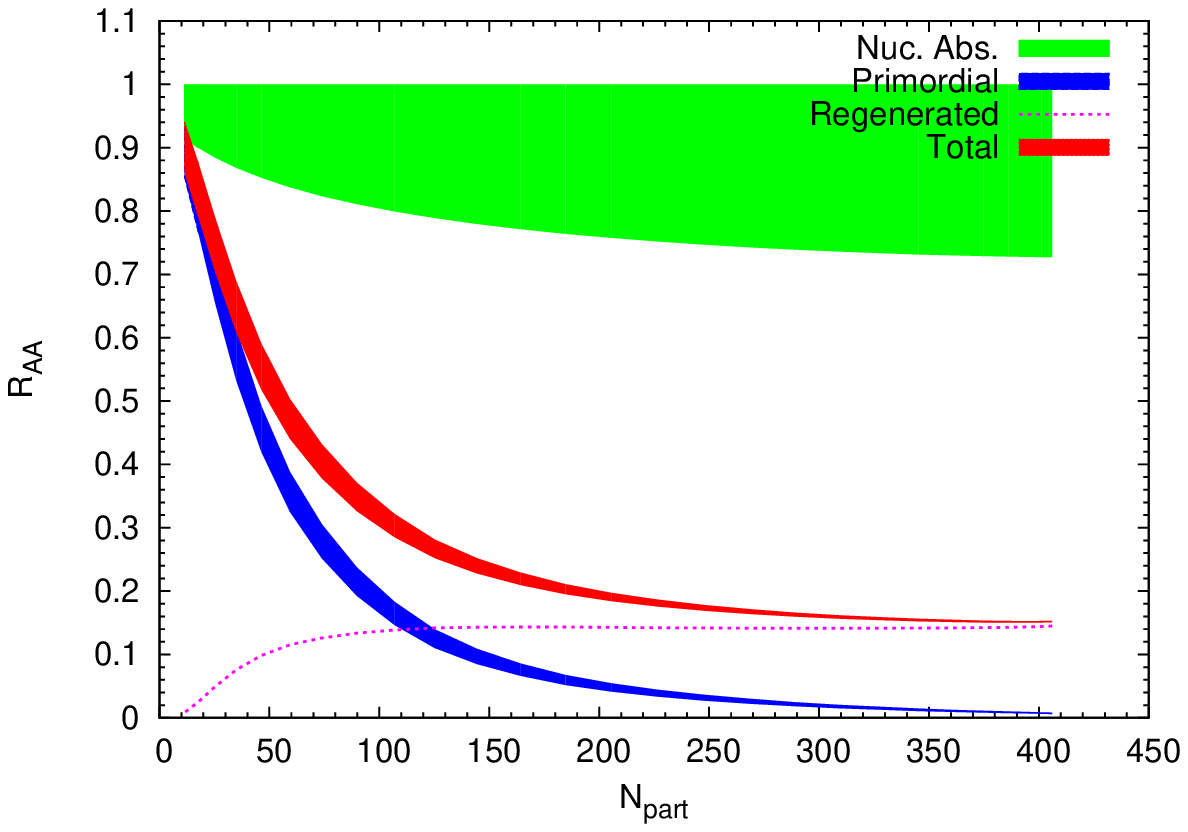}
\vspace{0.0cm}
\caption{The nuclear modification factor for inclusive $\Upsilon$ (top 
row, compared to CMS data~\cite{cms-11}), direct $\Upsilon$ (second row), 
$\Upsilon^{\prime}$ (third row) and $\chi_b$ (bottom row), as a function 
of centrality in Pb-Pb($\sqrt{s_{NN}}$=2.76\,TeV) collisions at LHC. 
The left column corresponds to the weak-binding scenario, the right one
to the strong-binding scenario.
In each panel, CNM effects alone are shown by the green band, CNM plus 
QGP suppression by the blue band, regeneration by the dashed pink line 
and the total by the red band.} 
\label{fig:lhc-raa}
 \end{figure*}

%%%%%%%%%%%%%%%%%%%%%%%%%%
Building upon previous $Y$ calculations for the LHC in Pb-Pb collisions
at $\sqrt{s_{NN}}$=5.5\,TeV~\cite{Grandchamp:2005yw} we examine 
the centrality dependence at $\sqrt{s_{NN}}$=2.76\,TeV, displayed
in Fig.~\ref{fig:lhc-raa} for inclusive and direct $\Upsilon$, as
well as for direct $\Upsilon'$ and $\chi_b$. 
The higher temperatures at LHC (relative to RHIC) reinforce the
stronger suppression in the weak-binding scenario compared to
strong binding, for inclusive and direct $\Upsilon$ production.
The CMS data~\cite{cms-11} are now significantly better described in 
the SBS, suggesting little effect of color-screening on the ground 
state even under LHC conditions. The initially (would-be) produced 
excited states exhibit close to complete suppression in both scenarios,
but significant regeneration occurs in the SBS, due to a much
larger equilibrium limit. This should be a measurable effect, especially
in connection with $p_t$ spectra (not discussed in our present work).
In addition, the feeddown of the regenerated excited states accounts 
for a significant portion of the observed (inclusive) $\Upsilon$(1S) 
regenerative component. Without feeddown, the regeneration effect is 
$\sim$5\%, while including feeddown it is roughly twice as large,
$\sim$10\%. The understanding of the behavior of the higher $\Upsilon'$ 
and $\chi_b$ states is thus essential to a proper interpretation of the 
observed $\Upsilon$ ground-state production at LHC. 

%%%%%%%%%%%%%%%%%%%%%%%%%%%%%%%%%%%%%%%%%%%%%%
\subsection{Comparison to Other Recent Work}
%%%%%%%%%%%%%%%%%%%%%%%%%%%%%%%%%%%%%%%%%%%%%
In this section we briefly discuss the results from other recent
calculations of $Y$ production in URHICs in comparison to our findings.

In Ref.~\cite{Liu:2010ej} $Y$ suppression at RHIC was studied using
gluo-dissociation with vacuum binding energies, folded over an ideal
hydrodynamic evolution. Dissociation temperatures were implemented
according to potential-model results based on $V$ or $U$ potentials.
In line with our results for the SBS, the $Y$(1S) remains essentially
unsuppressed in Au-Au($\sqrt{s_{NN}}$=0.2\,TeV) collisions, while
the yields and $p_T$-spectra of excited states show significant 
sensitivity to variations in the dissociation temperature. The larger 
dissociation temperatures
following from $V$=$U$ give results close to our SBS.

In Ref.~\cite{Strickland:2011mw} $Y$ suppression was computed using 
an in-medium Cornell potential~\cite{Karsch:1987pv} including a 
perturbative imaginary part and the effect of anisotropies. The 
imaginary part corresponds to quasifree 
dissociation~\cite{Grandchamp:2001pf,Riek:2010py}, and was folded over 
an ``extended anisotropic hydrodynamics" evolution. The resulting 
suppression in Pb-Pb at LHC is somewhat weaker than in our WBS, partly 
because the anisotropies decrease the $Y$ dissociation rates. 

In Ref.~\cite{Brezinski:2011ju} the gluo-dissociation cross section
was extended to approximately incorporate the string term in the 
$b\bar b$ potential. With in-medium binding energies but 
with a rather small coupling constant of $\alpha_s$=0.2, a schematic 
estimate of the $\Upsilon$ suppression in central Pb-Pb at LHC is 
roughly compatible with CMS data.

Close to completion of our work, Ref.~\cite{Song:2011nu} appeared where 
bottomonium production at RHIC and LHC is computed within the same 
rate-equation approach~\cite{Grandchamp:2003uw,Grandchamp:2005yw} 
as employed here, including regeneration.
The $Y$ dissociation widths were taken from a NLO perturbative QCD
calculation folded over a schematic viscous fireball evolution with an
ideal-gas equation of state. No constraints from lattice correlators 
on the $Y$ spectral functions were evaluated. Contrary to our results, 
a preference for an in-medium binding scenario has been inferred and no 
significant regeneration contributions are found even at LHC.

%%%%%%%%%%%%%%%%%%%%%%%%%
\section{Conclusions}
\label{sec:concl}
%%%%%%%%%%%%%%%%%%%%%%%%%
We have studied bottomonium production in ultrarelativistic heavy-ion
collisions at RHIC and LHC utilizing a kinetic rate-equation approach
in a quark-gluon plasma background, including suppression and 
regeneration mechanisms dictated by detailed balance. Our calculations 
are based on previous work in Ref.~\cite{Grandchamp:2005yw} where,
in particular, two scenarios for bottomonia in the QGP were 
evaluated, i.e. vacuum and in-medium bound-state properties (vis-a-vis
strong and weak-binding scenarios). At the time, the latter seemed to be 
the theoretically better motivated case. The present work, however, 
suggests a re-evaluation of this view. 
Firstly, the strong-binding limit provides a more natural explanation
of the near temperature independence of bottomonium correlators
now available from thermal lattice QCD. Furthermore, the application
to recent STAR and CMS data at RHIC and LHC (including updated 
open and hidden-bottom input cross sections) indicates that a more
stable $\Upsilon$ ground state gives a better description
of its nuclear modification factor. These findings corroborate the
picture that the $\Upsilon$ ground state is not much affected by
color screening up to fairly high temperatures of $\sim$3-4\,$T_c$.
We also found that regeneration
contributions, which have been neglected in most previous works, 
are not negligible, especially for strong binding at the LHC. 

Several lines of future studies emerge. The background medium needs to 
be improved using realistic hydrodynamic evolutions. Some work in this
direction has already been conducted. The regeneration term should be 
calculated with time-dependent bottom-quark spectra, constrained by 
$B$-meson observables. Based on those, transverse-momentum spectra and 
elliptic flow of bottomonia need to be calculated (again, pertinent
work has begun).
In all these, the consistency to the charmonium and open-charm sector
should be maintained. A comprehensive description of heavy-flavor
observables, based on theory constrained by lattice QCD, remains the
long-term goal, which appears to be achievable.

% % % % % % % % % % % % %
\acknowledgements 
% % % % % % % % % % % % % % %
This work was supported in part by the National Science Foundation (NSF) 
grant No. PHY-0969394 (RR), the Department of Energy grant no. 
DE-FG02-87ER40371 (XZ), the Texas A\&M Cyclotron Institute REU program
under NSF grant No. PHY-1004780 (AE), and by the Humboldt Foundation (RR).

%%%%%%%%%%%%%%%%%%%%%%%%%%%%%%%%%%%%%%%%%%%%%%%%

\end{document}